\documentclass[twocolumn,showpacs,preprintnumbers,amsmath,amssymb,pre]{revtex4-1}

\usepackage[T1]{fontenc}
\usepackage[utf8]{inputenc}
\usepackage{amstext}
\usepackage{amsmath}
\usepackage{amssymb}
\usepackage{graphicx}
\usepackage{latexsym}
\usepackage{esint}
\usepackage[english]{babel}
\usepackage{dcolumn}
\usepackage{bm}
\usepackage[FIGTOPCAP, hang, nooneline]{subfigure}
\usepackage[usenames,dvipsnames]{color} 
\usepackage{mathrsfs}
\usepackage{verbatim}
\usepackage{chemarr}
\usepackage[pdftex]{hyperref}
\usepackage{latexsym}
\usepackage[retainorgcmds]{IEEEtrantools}
\usepackage[usenames,dvipsnames]{color} 
\usepackage{amsbsy}
\usepackage{mathbbol}

\newcommand{\tpsi}{\bar\psi}

\newcommand{\acomment}[1]{#1}

\newcommand{\action}{{\mathcal{S}}}

\newcommand{\md}{{\mathrm{d}}}

\newcommand{\bz}{{\mathbf{z}}}
\newcommand{\bx}{{\mathbf{x}}}
\newcommand{\by}{{\mathbf{y}}}
\newcommand{\bq}{{\mathbf{q}}}
\newcommand{\bp}{{\mathbf{p}}}
\newcommand{\tphi}{{\bar{\phi}}}
\newcommand{\tchi}{\bar{\chi}}

\newcommand{\Tr}{\mbox{Tr}}

\newcommand{\trn}{\tau}
\newcommand{\ud}{\,\mathrm{d}}
\newcommand{\strich}{ }
\renewcommand{\kappa}{k}

\begin{document}

\title{Long-Range and Many-Body Effects in Coagulation Processes}

\author{Anton A. Winkler} \author{Erwin Frey}
\affiliation{Arnold Sommerfeld Center for Theoretical Physics and Center for NanoScience, Department~of~Physics, Ludwig-Maximilians-Universit\"at M\"unchen, Theresienstra{\ss}e 37, 80333 M\"unchen, Germany}

\begin{abstract}
We study the  problem of diffusing particles which coalesce upon contact. With the aid of a non-perturbative renormalization group, we first analyze the dynamics emerging below the critical dimension two, where strong fluctuations imply anomalously slow decay. 
Above two dimensions, the long-time, low-density behavior is known to conform with the law of mass action.
For this case, we establish an exact mapping between the physics at the microscopic scale (lattice structure, particle shape and size) and the macroscopic decay rate in the law of mass action. 
In addition, we identify a term violating this classical law. It originates in long-range and many-particle fluctuations and is a simple, universal function of the macroscopic decay rate. 
\end{abstract}

\acomment{\maketitle}

\section{Introduction}

The standard for the mathematical treatment of chemical reaction kinetics is provided by the law of mass action (LMA), stating that the rate of an elementary reaction is proportional to the product of the densities of its participants~\cite{Kuzovkov:1988p11649,Zhou:2010p16331}. However, unless  the environment is continually well-stirred, the 
interplay between reaction and dispersion implies fluctuations in space and time, which tend to promote retarded kinetics~\cite{redner_fractals,Frachebourg:1996p5056}. Indeed, typically there is a critical dimension below which ineffective diffusive mixing leads to a qualitative change in the speed of  the reaction, and one speaks of anomalous kinetics~\cite{Toussaint:1983p15598,Kang:1984p14320, *Kang:1984p11987, *Kang:1985p20980,Kopelman:1988p15594,Kuzovkov:1988p11649}. 
In contrast, above the critical dimension, one usually expects fluctuation only to alter the decay kinetics quantitively, by influencing the reaction rates~\cite{Zhou:2010p16331}. Yet, rigorous analysis often only supports the validity of the LMA for asymptotically long times and low densities, whereas analytical treatment away from this regime is rare~\cite{Doi:1976p15932,Mikhailov:1981p19705,*Mikhailov:1981p19688,*Gutin:1987p19618}. 
 
Employing a non-perturbative renormalization group (NPRG) approach, in this work  we explore a bimolecular reaction scheme where particles clot upon contact, $A+A \to A$, and explore space by diffusion. 
In the spirit of investigating idealized and simplified models to gain insight into microscopically much more complex processes, this scheme was introduced  by Smoluchowski to help understand the physics behind the coagulation of gold particles suspended in an electrolyte~\cite{Smoluchowski:1917p16756,Zsigmondy:1917p16758,Chandrasekhar:1943p17335}. 
This type of reactions is of interest for a variety of fields of natural science, with applications to the dynamics of aerosols~\cite{Lai:1972p23226}, the binding of proteins~\cite{Zhou:2010p16331}, the decay of defects in solids~\cite{Kopelman:1988p15594}, or the kinetics of chemical reactions~\cite{Waite:1957p23313,Kuzovkov:1988p11649}.

From perturbative renormalization group analysis it is known that, below the critical dimension two, the reaction kinetics displays critical behavior with anomalously slow decay of the density~\cite{Peliti:1986p11725,Lee:1994p10274}. We recover this result, and demonstrate that for the one-dimensional case the NPRG allows to obtain a substantially enhanced quantitative result as compared to the perturbative approach. 
The NPRG is also suited for the analysis above the critical dimension. 
We can thus show that for three-dimensional systems the LMA is violated by a relatively strong, non-analytic term in the density. The effects become more pronounced as the particle density increases and are attributed to long-range fluctuations and many-particle correlations. The corresponding additional term is a universal function of the non-universal, macroscopic decay rate (the proportionality factor in the LMA). 
To calculate the latter, we construct a mapping between the physics at the microscopic scale, defined by the shape and size of the particles and the structure of the lattice, and an effective macroscopic description.
To the best of our knowledge this is one of the very few instances where the renormalization group idea of mapping physics from the micro- to the macro-scale can be accomplished exactly. 

The rest of this article is organized as follows. In Section~\ref{sec:NPRG} we give a brief introduction to the NPRG formalism.  Starting from the microscopic definition of the process, it allows to calculate the \emph{effective action} $\Gamma$, by systematically integrating over the degrees of freedom, going from high to low momenta and frequencies. 
From $\Gamma$ the macroscopic behavior, such as the kinetic equation governing the time evolution of the density, can be deduced. 
After these preliminaries, we proceed in Section~\ref{sec:belowdc} to discuss the coagulation process below the critical dimension. 
We demonstrate that the specific properties of the problem at hand imply significant simplifications of the flow equation. This enables us to derive an adequate numerical description of the anomalous decay kinetics, characterized by universal parameters.
Finally, in Section~\ref{sec:abovedc} we turn to the treatment of the process above the critical dimension in a broader and more detailed discussion than is given in our previous paper on this topic~\cite{Winkler:2012p23673}.

\section{NPRG and Reaction-Diffusion Systems} 
 
\label{sec:NPRG} 

The NPRG based on the effective action $\Gamma$ is a suitable framework for the study of fluctuation effects in reaction-diffusion systems. Originally developed for the treatment of equilibrium physics~\cite{Wetterich:1993p9399,Ellwanger:1994p25598,Morris:1994p22097,Tetradis:1994p21967}, it has been adapted recently to non-equilibrium systems~\cite{Canet:2004p18293,Canet:2006p16432,Canet:2011p21015}. In the context of reaction-diffusion, it has led to new insights in the critical properties of branching and annihilating random walks and in the generalized voter class~\cite{Canet:2004p83,Canet:2004p18297,Canet:2005p4749}. 

For completeness and to set the notation we provide in this section a short introduction to the NPRG formalism. 
We specialize the treatment to the coagulation process and put particular emphasis on the derivation of the kinetic equation. 
For authoritative reviews and more detailed accounts on the general topic the reader is referred to~\cite{Berges:2002p16426,Canet:2004p18293,Canet:2006p16432,Canet:2011p21015,Delamotte:2012p25679}. 

\subsection{The Field-Theoretic Action}

\label{subsec:fieldtheoreticaction}

In a first step, the process is mapped onto a field theory by a Fock space formalism, an approach devised by several authors \cite{Doi:1976p168,Doi:1976p15932,Zeldovich:1978p19619,Grassberger:1980p21374,Peliti:1985p17344}, for a review see~\cite{Tauber:2005p493}. In our case of diffusing particles which coagulate upon contact, the action of the path integral can be split up as
\begin{equation}
\label{eq:actionannicoag}
	\action[\tphi,\phi] = \action_{Z}[\tphi,\phi] + \action_{\epsilon}[\tphi,\phi] + \action_{\lambda}[\tphi,\phi] \,.
\end{equation}
Here the fields $\phi = \phi(\bx,t) \in \mathbb{R}$ and $\tphi = \phi(\bx,t) \in i \mathbb{R}$ are related to particle annihilation and creation operators, respectively, and are  variables of the position $\bx$ on the lattice and of time $t$.   
There is a term for the time evolution
\begin{equation*}
	\action_Z[\tphi,\phi] = Z \sum_\bx \int \! \mathrm{d} {t} \,  \tphi(\bx,t) \partial_t \phi(\bx,t) \,,
\end{equation*}
where the factor $Z = 1$ (it may change along the renormalization group flow, although for the coagulation process this not the case, as discussed below).  
Furthermore, we have a diffusion term $\action_{\epsilon}[\tphi,\phi]$ (a particle can hop from site $\mathbf{x}$ to site $\mathbf{y}$ with rate $h$, when $\mathbf{x}$ and $\mathbf{y}$ are adjacent sites, indicated by brackets $< \bx, \by >$ in the following sum), which becomes
\begin{equation*}
	Z h \sum_{< \bx, \by >} \int \! \md t \, \left[\tphi(\bx,t) - \tphi(\by,t) \right]\big[ \phi(\bx,t) - \phi(\by,t) \big] 
\end{equation*}	
in position space (again $Z = 1$).
In Fourier space, due to translational invariance,  it is diagonal, 
\begin{equation}
		 \label{eq:epsq}
	\action_\epsilon[\tphi,\phi] = Z \int_{\mathbf{q},\omega} 
		 \epsilon(\mathbf{q}) \tphi(-\mathbf{q},-\omega) \phi(\mathbf{q},\omega)  \, ,
\end{equation}
where $\int_\omega := \int \! \frac{\mathrm{d} \omega}{2 \pi}$, and $\int_{\mathbf{q}} := \int \! \frac{\mathrm{d}^d q}{V_{B}}$ runs over the first Brillouin zone of volume $V_B$. Unless otherwise stated, in this article, we consider a hypercubic lattice, where  $V_B = \left(\frac{2 \pi}{a}\right)^d$ and the dispersion relation reads~\cite{Dupuis:2008p12392,*Machado:2010p11234}
\begin{equation*}
	\epsilon(\mathbf{q}) = 4 h \sum_{\nu = 1}^{d} \sin^2\!\left(\frac{q_\nu a}{2}\right) \,,
\end{equation*}
and we set the lattice spacing equal to one, $a = 1$, to define the length scale. In this case the hopping rate $h$ equals the diffusion constant $D$. To define the scale of time, we set, unless otherwise stated, $D = 1$.
 
Below the critical dimension, the  structure of the lattice does not influence the long-time, low-density behavior~\cite{Lee:1994p10274}. 
For the purposes of Section~\ref{sec:belowdc} it is therefore adequate to perform the continuum limit, where  $\epsilon(\mathbf{q}) =  q^2$, i.e. Eq.~(\ref{eq:epsq}) is replaced with
\begin{IEEEeqnarray*}{rCl}
	 \action_\epsilon[\tphi,\phi] &  = &
	 \int_{\mathbf{q},\omega} 	 q^2 \tphi(-\mathbf{q},-\omega) \phi(\mathbf{q},\omega) =  \\
	 & = & - \int \! \md \bx \, \md t  \, \tphi(\mathbf{x},t) \nabla^2 \phi(\mathbf{x},t)  \,.
\end{IEEEeqnarray*}

Finally in the action~(\ref{eq:actionannicoag}) there is a reaction term $\action_{\lambda}[\tphi,\phi]$ for coagulation $A +  A \to A$, which reads
\begin{equation*}
	\label{eq:actionlalapr}
	 \sum_{\bx,\by} \int \! \md t \,  \lambda(\by-\bx) \left[ \tphi(\by,t) + 1 \right] \tphi(\bx,t) \phi(\by,t) \phi(\bx,t)  \,.
\end{equation*}
In order to be able to study interactions with a finite range, we have introduced here the \emph{reaction kernel} $\lambda(\bz)$, which defines the shape and size of the particles. 
Thus, if there is a particle at site $\bx$, a particle at site $\by$ is annihilated with rate $\lambda(\by - \bx)$. 

Again,  below the critical dimension the microscopic properties are irrelevant for the universal behavior and it suffices to consider local interactions
\begin{equation*}
\action_{\lambda}[\tphi,\phi] = \int \! \md^d x \, \md t  \,  \lambda \left[ \tphi(\bx,t) + 1 \right] \tphi(\bx,t) \phi(\bx,t)^2 \,,
\end{equation*}
with $\lambda := \sum_{\bz} \lambda(\bz)$.

\subsection{The Wetterich Equation}

In analogy to equilibrium physics, for reaction-diffusion systems one can define a generating functional 
\begin{equation*}
	\mathcal{Z}[J, \bar J] \!= \!\int \mathcal{D} \bar \phi \mathcal{D} \phi \exp\!\left[ \! - \action[\tphi,\phi] + \sum_\bx \! \int \!  \mathrm{d}t\,  \left(\bar J \phi + J \bar \phi \right)\right]   \,,
\end{equation*}
from which observables such as the temporal evolution of the particle density can be derived. 
The fields $J, \bar J$ are introduced in order that the associated functional
\begin{equation*}
	W[J, \bar J ] = \ln \mathcal{Z}[J, \bar J] 
\end{equation*}
genererates the connected Green's functions by functional derivation in the fields $J$, $\bar J$ at $J=\bar J = 0$. 
We remark that the field $J$ induces particle input $\varnothing \to A$, as long as $J$ is positive. 

The degrees of freedom are integrated systematically, going from fast frequencies and short wavelengths at the microscopic scale $\Lambda$ (related to the typical reciprocal length scale of the interaction) to slow frequencies and long wavelengths. 
This is implemented by introducing a ``mass term'' $\Delta \action_{\kappa}$ to the action suppressing fluctuations of these modes,
\begin{IEEEeqnarray*}{rcl}
\label{eq:kdependentZ}
	 & \mathcal{Z}_{\kappa}[J, \bar J] = \int \mathcal{D} \bar \phi \mathcal{D} \phi \exp\Bigg[ - \action[\bar \phi, \phi] - \Delta \action_{\kappa}[\bar \phi, \phi] \: +  & \nonumber \\
	 & + \: \sum_\bx \int \! \md t \,  (\bar J \phi + J \bar \phi) \Bigg] \,, &
\end{IEEEeqnarray*}
where the mass term reads
\begin{equation*}
	\Delta \action_\kappa[\tphi,\phi] = \int_{\mathbf{q},\omega} R_\kappa(\bq) \tphi(-\mathbf{q},-\omega) \phi(\mathbf{q},\omega) \,,
\end{equation*}
with a cutoff function $R_\kappa$ that is independent of the frequency $\omega$, a convenient choice for reaction-diffusion processes~\cite{Canet:2004p18293}. 
Specifically, we take~\cite{Litim:2001p13475, Dupuis:2008p12392,*Machado:2010p11234} 
\begin{equation}
\label{eq:dupuis}
	R_\kappa(\mathbf{q}) = \left(\kappa^2 - \epsilon(\mathbf{q})\right) \, \Theta (\kappa^2 - \epsilon(\mathbf{q})) \,.
\end{equation}

Instead of calculating the renormalization group flow of the functional $W_\kappa[J, \bar J ] = \ln \mathcal{Z}_\kappa[J, \bar J]$~\cite{Wegner:1973p21527,Polchinski:1984p21517}, for the approach employed in this article one considers
the so called effective average action $\Gamma_\kappa$. 
Denoting the expectation values
\begin{equation}
	\label{eq:Jpsi}
	\tpsi := \langle \tphi \rangle = \delta W_{\kappa}/\delta         J \,, \quad
	\psi := \langle \phi \rangle   = \delta W_{\kappa}/\delta \bar J       \,, 
\end{equation}
at $J = \bar J = 0$, it is defined by
\begin{equation}
\label{eq:Gklegendre}
	\Gamma_{\kappa}[\tpsi,\psi] = \sum_{\bx} \int \!  \md t \,  \left(\bar J \psi  + J \tpsi\right) - \Delta \action_{\kappa}[\tpsi,\psi] - W_{\kappa}[J,\bar J] \,.
\end{equation}
Notice that this is just the Legendre transform of $W_\kappa$, up to the term  $\Delta \action_{\kappa}$, which is added for mathematical convenience and vanishes as $\kappa \to 0$.

At the microscopic scale $\Lambda$ the cutoff function $\Delta \action_\kappa$  is supposed to be large, so that it freezes the deviations from the expectation values $\psi$, and $\tpsi$, and  renders the functional trivial. The initial condition thus becomes (see~\cite{Canet:2011p21015} for a careful discussion)
\begin{equation*}
	\label{eq:wettinitialcond}
	\Gamma_{\kappa=\Lambda}[\tpsi,\psi] = \action[\tpsi,\psi] \,.
\end{equation*}

Only modes with $q \lesssim \Lambda$ are integrated out along the renormalization group flow. Therefore, if we employ the continuum limit in the action $\action$ (as in Section~\ref{sec:belowdc}), a finite $\Lambda$ sets the ultraviolet cutoff. If otherwise we keep the discrete lattice structure, $\mathbf{q}$ only runs over the first Brillouin zone, intrinsically enforcing an ultraviolet cutoff. Similarly, we have an implicit ultraviolet cutoff for finite-size objects, such as balls with radius $R$, where modes with $q \gg R^{-1}$ are suppressed. In these cases there is no need for an explicit cutoff and we can set $\Lambda = \infty$. 

The central equation of the NPRG formalism is  the Wetterich equation. This flow equation connects the effective average action at the microscopic scale $\Gamma_{\kappa=\Lambda} = \action$, where fluctuations are neglected, with the effective action $\Gamma = \Gamma_{\kappa=0}$ where all degrees of freedom are integrated.
It can be expressed as 
\begin{equation}
	\label{eq:wetter}
	\partial_\kappa \Gamma_\kappa[\tpsi,\psi] = \frac{1}{2} \Tr \left[ \partial_{\kappa} \hat R_{\kappa} \left(\hat \Gamma_{\kappa}^{(2)}[\tpsi,\psi] + \hat R_{\kappa} \right)^{-1} \right]  \,,
\end{equation}
where $\hat \Gamma_k^{(2)}$ and $\hat R_k$ denote the $2 \times 2$ matrices of the second functional derivatives of $\Gamma_k$ and of the mass term $\Delta S_k$, respectively, and $\Tr$ denotes the trace.

\subsection{The Derivative Expansion}

In general, the Wetterich equation cannot be solved exactly.
The derivative expansion is an approximation based on the fact that 
infrared singularities are suppressed by the cutoff function $R_{\kappa}$ and therefore the effective average action $\Gamma_{\kappa}[\tpsi,\psi]$ is analytic so long as the scale $\kappa >0$~\cite{Berges:2002p16426}. It is performed by expanding the functional $\Gamma_{\kappa}[\tpsi,\psi]$ in orders of the temporal derivative $\partial_t$ and the spatial derivative $\nabla$,  
truncating $\Gamma_{\kappa}[\tpsi,\psi]$ at a certain order. 

For the following expansions, let us take the continuum limit of the action as the initial condition. 
A common truncation of the effective average action is the ``leading order'' approximation 
\begin{equation}
	\Gamma_{\kappa}[\tpsi,\psi] = \int \! \mathrm{d}^d x\, \mathrm{d}t \left[U_{\kappa}(\tpsi,\psi) + Z_{\kappa} \tpsi( \partial_t - D_{\kappa} \nabla^2) \psi \right]\,. 
\label{eq:dominantorder}
\end{equation}
This approximation is popular because it already allows to determine the anomalous dimension $\eta = - \kappa\, \partial_{\kappa} \ln Z_{\kappa}$ and the dynamic exponent $z = 2 + \kappa \partial_{\kappa} \ln D_{\kappa}$. However, we show below that for the coagulation process this ansatz is equivalent to a lower order in the approximation, the so called ``local potential'' approximation,
\begin{equation}
\label{local_potential_approx}
	\Gamma_{\kappa}[\tpsi,\psi] = \int \! \mathrm{d}^d x \, \mathrm{d}t \left[U_{\kappa}(\tpsi,\psi) + \tpsi\left(\partial_t - \nabla^2\right) \psi \right]\,,
\end{equation}
where only the renormalization of the local potential $U_{\kappa}$ is taken into account. The reason is that, as proven in Section~\ref{sec:belowdc}, for our particular process $Z_{\kappa}$ and $D_{\kappa}$ are not affected by the renormalization group flow, i.e.~$Z_{\kappa} = D_{\kappa} = 1$ for all $\kappa$.  Thus, for the coagulation process $\eta$ and $z$  agree with the mean-field exponents 0 and 2, respectively. 

Let us consider homogeneous fields $\tpsi(x,t) \equiv \tpsi,\,\psi(x,t) \equiv \psi$.  The relation between the average effective action and the local potential then reads
\begin{equation*}
	\Gamma_{\kappa}[\tpsi,\psi] = V T \cdot U_{\kappa}(\tpsi,\psi)\,,
\label{eq:effective_local}
\end{equation*}
with the asymptotically large volumes of space and time $V = \int \! \md^d x = (2\pi)^d \delta(\mathbf{p} = \mathbf{0})$ and $T = \int \! \mathrm{d}t = 2\pi \delta(\omega= 0)$, respectively. With the cutoff function~(\ref{eq:dupuis}) (where $\epsilon(\bq) = q^2$, since we consider the continuum limit),  from the Wetterich equation~(\ref{eq:wetter}) one can deduce the flow equation for the effective average potential, 
\begin{equation}
\label{dUk}
	\partial_{\kappa} U_{\kappa} = \frac{\widetilde{V}_d \kappa^{d+1} \left(U_{\kappa}^{(1,1)} + \kappa^2\right)}{ \sqrt{\left(U_{\kappa}^{(1,1)} +  \kappa^2\right)^2  - U_{\kappa}^{(2,0)} U_{\kappa}^{(0,2)}}}\,.
\end{equation}
Here  $\widetilde{V}_d = \frac{V_d}{(2 \pi)^d}$, $V_d$ denotes the volume of the $d$-dimensional unit sphere, and 
$U_{\kappa}^{(m,n)} = \frac{\partial^{m+n} U_\kappa}{\partial \tpsi^m \partial \psi^n}$ the $m$th and $n$th derivative of $U_\kappa$ with respect to $\tpsi$ and $\psi$, respectively.
In our case the initial condition reads
\begin{equation*}
	\label{dutinitial}
	U_{\Lambda}(\tpsi,\psi) = \lambda \tpsi^2 \psi^2 + \lambda  \tpsi \psi^2 \,.
\end{equation*}

\subsection{Dimensionless Flow Equation}

In order to be able to study critical behavior, we need to resolve the fixed points of the flow. 
As verified in the next section, this is realized if we introduce the dimensionless coordinates
\begin{equation}
\label{eq:rescaledcoord}
	\mathbf{x} = \kappa^{-1} \tilde{\mathbf{x}} \,, \quad t = \kappa^{-2} \tilde{t} \,, \quad  \mathbf{q} = \kappa \tilde{\bq} \,, \quad \omega = \kappa^{2} \tilde{\omega} \,,
\end{equation}
and the renormalized dimensionless fields 
\begin{equation}
\label{eq:rescaledfields}
	\tpsi(\mathbf{x},t) = \tchi(\tilde{\mathbf{x}},\tilde{t})  \,,\quad  \psi(\mathbf{x},t) = \kappa^d  \chi(\tilde{\mathbf{x}},\tilde{t}) \,.
\end{equation}	
Also, we introduce the renormalization time $\tau = \ln(\kappa/\Lambda)$. 
The proper dimensionless and renormalized form of the local potential obeys
\begin{equation*}
	U_{\kappa}(\tpsi,\psi) = \kappa^{d+2}  u_{\tau}(\tchi,\chi) \,.
	\label{eq:dimlesspot}
\end{equation*}
The cutoff function~(\ref{eq:dupuis}) now takes the form $R_k(q) = q^2 r(\tilde{q}^2)$ with $\tau$-independent
\begin{equation}
	r(\tilde{q}^2) = \left( \frac{1}{\tilde{q}^2} - 1 \right) \Theta\left(1-\tilde{q}^2\right) \,.
\label{eq:dimlesslitim}
\end{equation}
The flow for the dimensionless potential follows from Eq.~(\ref{dUk}), 
\begin{IEEEeqnarray}{rCl}
\partial_\tau u_{\tau} & = & - (d+2) u_{\tau} + 
	d \chi \, u_{\tau}^{(0,1)}  \: + \smallskip \nonumber \\
	& & + \:  
	  \frac{{\widetilde{V}_d} \left( u_{\tau}^{(1,1)} + 1 \right)}{ \sqrt{\left(u_{\tau}^{(1,1)} + 1\right)^2  - u_{\tau}^{(2,0)} u_{\tau}^{(0,2)}  }} \,,
\label{eq:dut}
\end{IEEEeqnarray}
with the initial condition
\begin{equation*}
	\label{dutinitial}
	u_{\tau=0}(\tchi,\chi) = \tilde\lambda_{\tau=0} \tchi^2 \chi^2 + \tilde\lambda_{\tau=0}  \tchi \chi^2 \,,
\end{equation*}
where $\tilde \lambda_{\tau = 0} = \Lambda^{d-2} \lambda$ is the rescaled coagulation rate.

\subsection{The Kinetic Equation}

As compared to the alternative non-perturbative approach of~\cite{Wegner:1973p21527,Polchinski:1984p21517} an advantage in calculating the effective average action $\Gamma[\tpsi,\psi]$ instead of $W[J,\bar J] = \ln \mathcal{Z}[J, \bar J]$  
is that it allows direct access to the observable $\psi$, the average density of the particles. At $\kappa = 0$, when all degrees of freedom are integrated out, the cutoff term vanishes, $\Delta \action_{\kappa=0} = 0$,  and the effective average action equals the effective action, $\Gamma_{\kappa=0} = \Gamma$. As a consequence of Eqs.~(\ref{eq:Jpsi},\ref{eq:Gklegendre}), the equation of motion is obtained by
the ``\emph{extremal principle}'' 
\begin{equation}
\label{eq:ofmotion}
	\delta \Gamma / \delta \psi = 0\,, \quad   \delta \Gamma / \delta \tpsi = J \quad \text{at } \tpsi = 0,\,\psi = \rho \,.
\end{equation}
Here we have allowed for particle input with rate $J(\bx,t) \ge 0$. The first equation is fulfilled by $\tpsi = 0$. Identifying $\psi$ with the particle density $\rho$, the second equation then determines the kinetics of the process. 

The extremal principle is the macroscopic analog of the classical field equations, $\delta \action / \delta \psi = 0$, $\delta \action / \delta \tpsi = J$, 
which give the mean-field equation for the particle density. Notice that in contrast to the ``microscopic'' action~$\action$, in the effective action~$\Gamma$, fluctuations in the particle numbers and correlations in space and time are taken into account.

When $\kappa > 0$, the effective average action $\Gamma_{\kappa}$ can be written in an expansion in the fields $\tpsi$, $\psi$, and multiple derivatives in space and time thereof \cite{Berges:2002p16426}.  For the purposes of this article, it is sufficient to restrict the time evolution to homogeneous fields, i.e.~$\psi(\mathbf{x},t) \equiv \psi(t)$. This discards derivatives in space. Neglecting also derivatives in time except for the term $\tpsi \partial_t \psi$, which is already present in the initial action, the general form of the effective average action reads
\[
	\Gamma_{\kappa}[\tpsi,\psi]  =  \sum_{\mathbf{x}} \int  \! \mathrm{d}t \, U_{\kappa}(\tpsi,\psi) +\sum_{\mathbf{x}} \int  \! \mathrm{d}t \,  \tpsi \partial_t \psi \,, 
\]
with the local potential $U_{\kappa}$, which can be defined for constant fields $\tpsi,\psi$ by $\Gamma_{\kappa}[\tpsi,\psi] = V T \, U_{\kappa}(\tpsi,\psi)$, and with the volumes of space and time $V = \sum_\bx$ (for a hypercubic lattice with unit lattice spacing), $T = \int  \! \mathrm{d}t$. The extremal principle~(\ref{eq:ofmotion}) then yields  
\begin{equation*}
	0 = \left.\frac{\delta \Gamma[\tpsi,\rho]}{ \delta \tpsi(\bx,t)}\right|_{\tpsi=0} = \partial_{\tpsi} U_{\kappa=0}(\tpsi,\rho)|_{\tpsi = 0} + \partial_t \rho \,.
\end{equation*}
With the \emph{non-equilibrium force} 
\begin{equation*}
	F(\rho) =  \partial_{\tpsi} U_{\kappa=0}(\tpsi,\rho)|_{\tpsi = 0} 
	\label{eq:nonequilifor}
\end{equation*}
this gives the kinetic equation
\begin{equation}
	\partial_t \rho(t) = - F(\rho(t)) \,.
	\label{eq:genkin}
\end{equation}

The non-equilibrium force $F$ can be determined in the simulations by introducing  homogeneous particle input with rate $J$. The extremal principle then implies $\partial_t \rho =  -F(\rho) + J$, such that for stationary states the non-equilibrium force equals the input rate, $F(\rho) = J$.

\section{Renormalization Below the Critical Dimension}

\label{sec:belowdc}

Although the coagulation process does not display a phase transition, its long-time approach to a vacant system can be described within the framework of critical phenomena.
Indeed, the process is suitable for treatment with the perturbative renormalization group approach,  
as was demonstrated in the pioneering works of Peliti~\cite{Peliti:1986p11725} and Lee~\cite{Lee:1994p10274}. 
Peliti established that the process displays an upper critical dimension and that its value is $d_c = 2$, confirming predictions based on heuristic arguments and  computer simulations \cite{Toussaint:1983p15598,Meakin:1984p12049,Kang:1984p14320,*Kang:1984p11987,*Kang:1985p20980}. 
Below  this critical dimension 
the density decay $\rho \sim \mathcal{A}_d (D t)^{-\frac{d}{2}}$ (for some dimension dependent amplitude $\mathcal{A}_d$ and diffusion constant $D$) is significantly retarded by fluctuations as compared to the classical behavior $\rho \sim \mathcal{A} (D t)^{-1}$.
Lee showed that the decay amplitude $\mathcal{A}_d$ is amenable to perturbative renormalization group analysis  near the critical dimension,
\begin{equation*}
\mathcal{A}_d =  
\begin{cases}
 	\frac{\ln(t)}{4 \pi} & \text{if } d=d_c =2\,, \\ 
  	\frac{1}{2 \pi \epsilon} + \frac{2 \ln(8 \pi) - 5}{16 \pi} + O(\epsilon) & \text{if } d< d_c \,,
\end{cases}
	\label{eq:lee}
\end{equation*}
where $\epsilon = 2 - d$ is the difference between the upper critical dimension $d_c = 2$ and the dimension $d$. 

In this section, we first study the mathematical properties of the effective average action $\Gamma_\kappa$, exploiting special properties of the coagulation process. 
Similar to restrictions due to certain symmetries in, say, magnetic models, we show that in the Taylor expansion of the effective average potential $U_\kappa$ many terms are not generated along the renormalization group flow. This can most conveniently be seen upon representing the flow by one-loop Feynman diagrams. 
We then proceed to exploit the relative simplicity of the flow equations for the study of the coagulation process in one dimension. 
The calculation cannot be carried out exactly and we need to recur to an approximation scheme, as introduced in the previous section, in order to reduce the complexity of the flow equation. However, the coagulation process permits us to go to a relatively high order of the approximation, therefore promising accurate results. Indeed, they compare well to the exact solution for one dimension. We also extend our analysis to general dimension $d\le2$ and thus reproduce results from  perturbative calculations for small $\epsilon$.

\subsection{The One-Loop Expansion and Restrictions on the Flow Equation}

Symmetries which are obeyed by the effective average action $\Gamma_{\kappa}$ usually play a central role in the discussion of the critical behavior of a system. Examples include the $O(N)$ symmetries in equilibrium statistical mechanics~\cite{Berges:2002p16426}, and, in non-equilibrium systems, the KPZ-symmetries~\cite{Canet:2010p18250,Canet:2011p23618}, the time-reversal symmetries of ``model A''~\cite{Canet:2007p9386,Canet:2011p21015}, or the so called ``rapidity symmetry'' of the contact process~\cite{Canet:2004p18293,Canet:2006p16432}. 
Similarly,  one can make statements on the mathematical properties of the effective average action $\Gamma_\kappa$ for the coagulation process. They do not come as an invariance with respect to a symmetry transformation of the fields $\tpsi$ and $\psi$ (or $\tchi$ and $\chi$).  
Rather, they become apparent upon expanding the effective average action $\Gamma_\kappa$, which is rendered analytic by the infrared cutoff as long as the scale $\kappa>0$~\cite{Berges:2002p16426}. 

Our goal in this section is to calculate  the dimensionless effective average potential $u_\tau(\tchi,\chi) = \kappa^{-d-2} U_{\kappa}(\tpsi,\psi)$. For a general reaction-diffusion process, it can be expressed as a power series~\cite{Canet:2011p21015}
\begin{equation*}
	u_\tau(\tchi,\chi) = \sum_{m \ge 1,n \ge 1}\frac{1}{m! n!}  \tilde{g}_{\tau}^{(m,n)} \tchi^m \chi^n\,.
\end{equation*}
For the concrete calculations later in this section, we employ the flow equation~(\ref{eq:dut}). 
To observe the specific properties of the potential $u_\tau$, however, 
it is favorable to revert to the full Wetterich equation~(\ref{eq:wetter}).  It can be recast in a form that allows for a diagrammatic analysis and thus reveals the mathematical structure more immediately,
\begin{equation}
\label{eq:one_loop}
	\partial_{\kappa} \Gamma_{\kappa} [\tpsi,\psi] =  \tilde{\partial}_{\kappa}  \underbrace{\frac{1}{2}\Tr \left[\ln\left( \hat \Gamma_{\kappa}^{(2)} [\tpsi,\psi] + \hat R_{\kappa} \right)\right]}_{\mathcal{D}_{\kappa}}    \,,
\end{equation}
where the derivative  $\tilde{\partial}_{\kappa} := \partial_{\kappa} R_{\kappa} \cdot \partial_{R_{\kappa}}$ acts only on the $\kappa$-dependence of the cutoff function $R_{\kappa}$. 

The functional $\mathcal{D}_{\kappa}$ on the right hand side in  Eq.~(\ref{eq:one_loop}) is known from perturbative analysis as the generator of one-loop Feynman diagrams~\cite{ZinnJustin,Berges:2002p16426}.
Therefore, the renormalization group flow of the $(m,n)$-point vertex functions $\Gamma^{(m,n)}_k$
(obtained by taking $m$ and $n$ functional derivatives of $\Gamma_k$ with respect to $\tpsi$ and $\psi$, respectively, at zero fields $\tpsi = \psi = 0$) 
can be represented by the one-loop Feynman diagrams for the $(m,n)$-vertex.
In this way, we also obtain the flow of the coefficients $\tilde{g}_\tau^{(m,n)}$, which are just proportional to 
$\Gamma^{(m,n)}_{\kappa\, (\mathbf{p}_1,\omega_1;\ldots;\mathbf{p}_{m+n},\omega_{m+n})}$ (where the functional derivatives are taken with respect to $\tpsi(\mathbf{p}_i,\omega_i)$, $i \in \{1,\ldots, m\}$ and $\psi(\mathbf{p}_{j},\omega_{j})$, $j\in \{m+1,\ldots, m+n \}$) at zero momenta and frequencies. 

Let us consider the flow of a general $(m,n)$-point vertex function, determined by the sum over the corresponding one-loop diagrams.  Due to causality, the propagator  only connects earlier to later vertices. This fact drastically restricts the number of possible Feynman diagrams. Since in the initial action the number $n$ of incoming legs is larger than the number $m$ of outgoing legs for all non-zero vertex functions, it is impossible to construct one-loop diagrams with $m > n$ from these vertices. Therefore the flow of the vertex functions with $m > n$ is zero. Similarly, the minimum number of incoming legs in the Feynman diagrams, which is $n = 2$, and the minimum number of outgoing legs,  which is $m=1$, is inherited to all scales $\kappa$. 

Now, from the fact that the number of legs cannot increase along the time arrow, it is readily deduced that the flow of $\Gamma_k^{(1,1)}$ vanishes, since there is no diagram of the form of Fig.~\ref{fig:pre_graphs}(a). 
\begin{figure}[]
\label{fig:pre_graphs}
      \centering \includegraphics[width=0.5\textwidth]{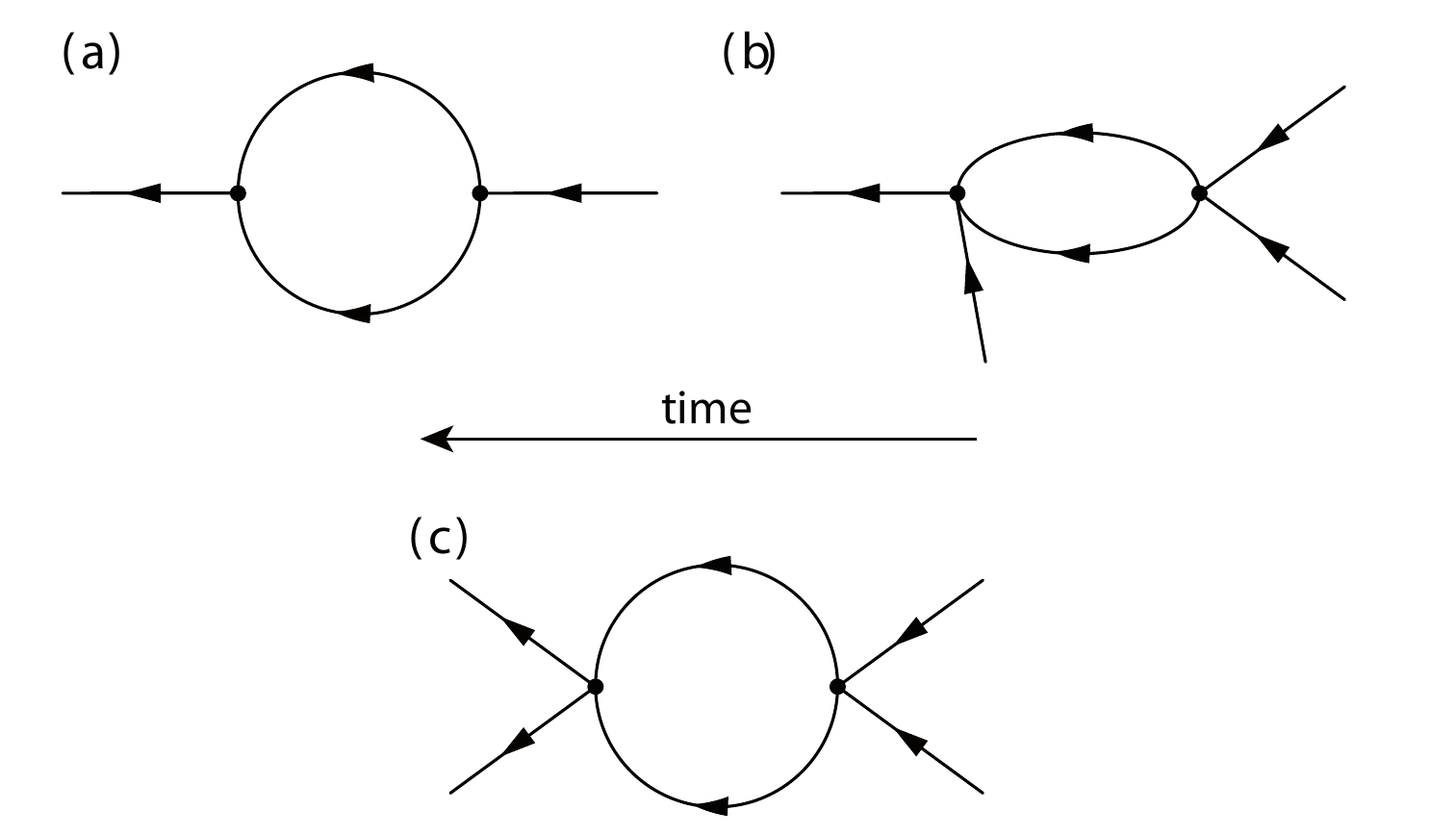}
      \caption{Important one-loop diagrams determining the flow of the vertex functions $\Gamma_\kappa^{(m,n)}$, where $m$ and $n$ are the number of outgoing and incoming legs, respectively. The fact the number of legs can only decrease along the time arrow puts significant restrictions on the possible diagrams. Diagram $(a)$ is not created in the flow equation because the (2,1)-vertex function is always zero. It follows that the propagator is not renormalized. Diagram~$(b)$ implies a term linear in $\Gamma_\kappa^{(1,3)}$ to the flow of $\Gamma_\kappa^{(1,3)}$. Similarly there arise linear contributions to the flow of general $\Gamma_\kappa^{(m,n)}$ (except for $m=n=2$, where there is a quadratic term in $\Gamma_\kappa^{(2,2)}$). Since diagram~$(c)$ is composed  only of (2,2)-vertices and propagators, the flow  for $\Gamma_\kappa^{(2,2)}$ can be solved without knowledge of further vertex functions.}  
\label{fig:pre_graphs}
\end{figure}
As a consequence, similar to the absence of propagator renormalization in perturbative renormalization~\cite{Peliti:1986p11725}, the factors $D_{\kappa}$ and  $Z_{\kappa}$ are constant
\begin{equation*}
	D_{\kappa} = Z_{\kappa} = 1\,.
\end{equation*}
Therefore, the leading order approximation~(\ref{eq:dominantorder}) and the local potential approximation~(\ref{local_potential_approx}) are equivalent for the coagulation process. 

Not only can we rule out certain vertex functions, but we can also make statements on the functional dependence of $\Gamma^{(m,n)}_k$ on other vertex functions. 
We first note that the one-loop diagrams for the flow of an $(m,n)$-vertex function evidently must not contain $(m^\prime,n^\prime)$-vertices with $n^\prime - m^\prime > n - m$ (the number of legs at one vertex in the diagram cannot decrease by more than the overall decrease $n-m$ in the number of legs). 

Moreover,  for $m \le n$ (except for $m=n=2$),  the flow $\partial_{\kappa} \Gamma^{(m,n)}_{\kappa}$ is linear in  $\Gamma^{(m,n)}_{\kappa}$: For $m < n$ the corresponding vertex has only $m$ outgoing lines, which cannot connect again to a vertex with $n$ incoming lines. For $m=n > 2$ one loop would not suffice to include a second $(m,n)$-vertex in the diagram. 
In general, one-loop diagrams for the flow of the vertex function $\Gamma^{(m,n)}_{\kappa}$ which contain one $(m,n)$-vertex can only involve exactly one additional vertex, which must be a $(2,2)$-vertex. 
(For the purpose of illustration, in Fig.~\ref{fig:pre_graphs}(b),  a  one-loop diagram to the $(1,3)$-vertex is shown.) 
Therefore, this linear term does not vanish.  

\begin{figure}[tb]
	\centering
	\mbox{\includegraphics[width=0.42\textwidth]{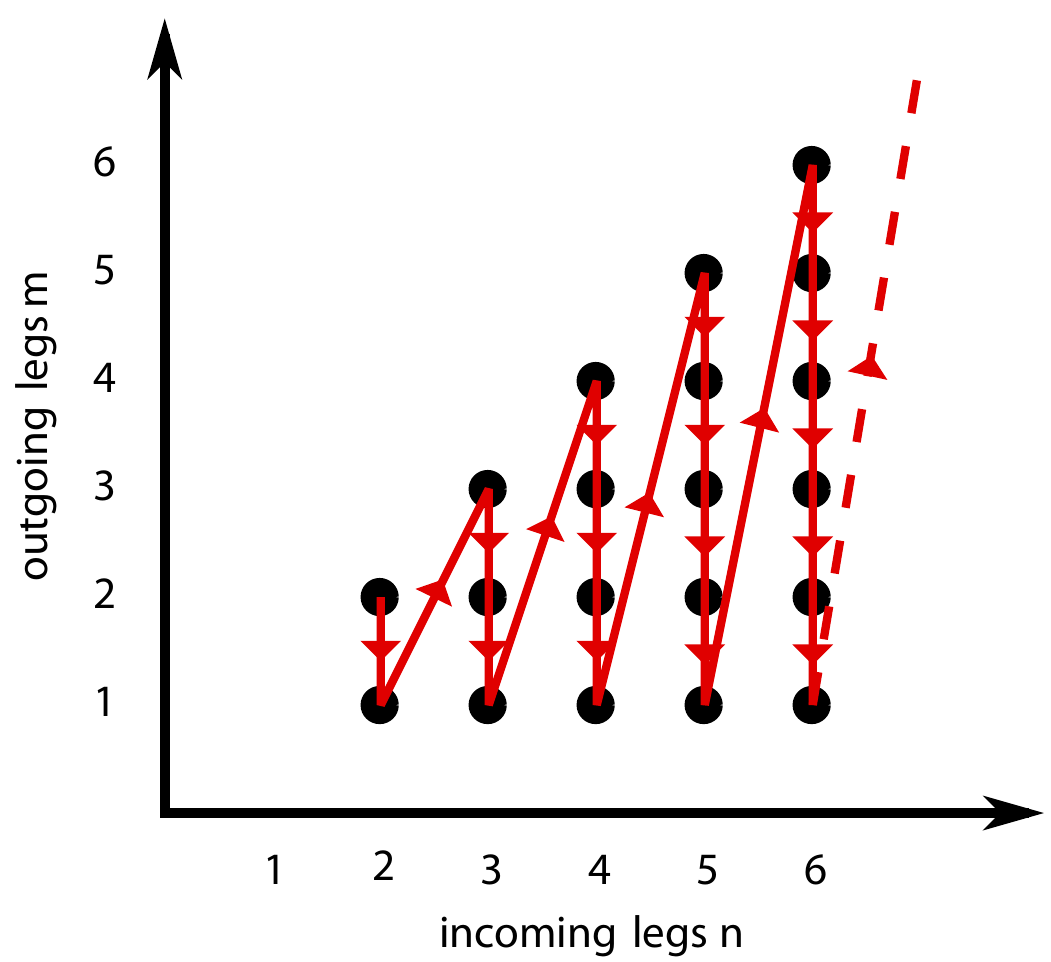}}
\caption{Illustration of the order for calculating the non-zero fixed-point coefficients $\tilde{g}^{\star(m,n)}$. 
In each step the result is independent of the coefficients to follow. This can be shown by looking at all possible one-loop diagrams which determine the flow of the vertex function $\Gamma^{(m,n)}_\kappa$ with $n$ incoming and $m$ outgoing legs.
For the physically relevant line with $m = 1$ all the coefficients $\tilde{g}^{\star(m^\prime,n^\prime)}$ with $n^\prime \le n$ must be known. In contrast, the diagonal elements $g^{\star(n,n)}$ only depend on $g^{\star(n^\prime,n^\prime)}$ with $n^\prime < n$.
}
\label{fig:reihenfolge}
\end{figure}

Let us apply our findings to the dimensionless  potential $u_\trn(\tchi,\chi)$ with the  
Taylor expansion
\begin{equation*}
	u_\trn(\tchi,\chi) = \sum_{m \ge 1, n \ge 2, m \le n} \frac{1}{m! n!}  \tilde{g}_\trn^{(m,n)} \tchi^m \chi^n \,.
\end{equation*}
Within the local potential approximation we have that  $\Gamma^{(m,n)}_{\kappa\,(\mathbf{p}_1,\omega_1;\ldots;\mathbf{p}_{m+n},\omega_{m+n})} = \Gamma^{(m,n)}_{\kappa\,(\mathbf{0},0,\ldots,\mathbf{0},0)}  \propto \tilde{g}_{\tau}^{(m,n)} $  ($\sum_i \bp_i = 0, \sum_i \omega_i = 0$). 
Thus, we can calculate the coefficients $\tilde{g}^{\star(m,n)}$ of the fixed-point potential $u^\star$  
within the local potential approximation step by step as follows (also see Fig.~\ref{fig:reihenfolge}): 
We start with $\tilde{g}^{\star(2,2)}$, which is easily obtained because $\partial_\trn \tilde{g}_\trn^{(2,2)}$  depends only on $\tilde{g}_\trn^{(2,2)}$, cf.~Fig.~\ref{fig:pre_graphs}(c).  We  then turn to $\tilde{g}^{\star(1,2)}$, whose flow $\partial_\trn \tilde{g}^{(1,2)}_\tau$  is a function of $\tilde{g}_\trn^{(2,2)}$ and $\tilde{g}_\trn^{(1,2)}$. Assuming that we know the fixed-point values of $\tilde{g}^{\star(m,n)}$ for all $m < n$ we can go on to treat $\tilde{g}^{\star(m,n)}$  successively for $m=n,n-1,\ldots,1$. In each step one simply needs to solve the linear equation 
\begin{equation*}
0  = c_1(m,n) + c_2(m,n) \cdot \tilde{g}^{\star(m,n)} \,,
\label{eq:lin}
\end{equation*}
given some $c_1(m,n)$ and $c_2(m,n) \ne 0$. More precisely, since $c_2(m,n) = c^\prime(m,n) \tilde{g}_\trn^{(2,2)} \tilde{g}_{\tau}^{(m,n)}$ with positive $c^\prime(m,n)$ and, as we show below,   positive $\tilde{g}_\trn^{(2,2)}$, we have that  $c_2(m,n) > 0$ below the critical dimension.

\subsection{The Fixed Point and the Upper Critical Dimension}
\label{subsec:critdim}

Let us have a closer look at the couplings for two-particle interaction,  $\frac{1}{4} \tilde{g}_\trn^{(2,2)} \equiv \frac{1}{2} \tilde{g}_\trn^{(1,2)} =: \tilde{\lambda}_\trn$. From the flow equation for the rescaled potential, Eq.~(\ref{eq:dut}), we obtain
\begin{equation*}	
	\partial_\trn \tilde{\lambda}_\trn 
	= (d-2) \tilde{\lambda}_\trn + 2 \widetilde{V}_d \tilde{\lambda}_\trn^2\,.
\label{eq:dtlambda}
\end{equation*}	
At the upper critical dimension $d_c = 2$ there is a transcritical bifurcation, such that, when the dimension $d < 2$, there is an unstable fixed point at $\tilde\lambda_\tau = 0$ (recall that $\tau$ flows in the negative direction) and a stable one at 
\begin{equation}
	\tilde\lambda_\tau = \tilde{\lambda}^\star = \frac{2-d}{2 \widetilde{V}_d}\,.
	\label{lambda_fixed}
\end{equation}
For $d>2$ the stability of the fixed points is interchanged and  at $d = 2 $ they merge to one, marginally stable fixed point. 
 
Similar behavior is observed for the other rescaled coefficients $\tilde{g}_\trn^{(m,n)}$. Once the coefficients $\tilde{g}_\trn^{(m^\prime,n^\prime)}$ which precede  $\tilde{g}_\trn^{(m,n)}$ in the order of the calculation indicated in Fig.~\ref{fig:reihenfolge} have relaxed to their fixed point $\tilde{g}^{\star(m^\prime,n^\prime)}$, their flow is described by an equation of the form 
\begin{equation*}
	\partial_\trn \tilde{g}_\trn^{(m,n)} = c_1(m,n) + c_2(m,n) \cdot \tilde{g}_\trn^{(m,n)}  \,,
	\label{eq:coefflineq}
\end{equation*}
with strictly positive $c_2(n,m)$ when $d < 2$. Hence $g_\trn^{(m,n)}$ approaches the finite and stable fixed point $\tilde{g}^{\star(m,n)} = -c_1(m,n)/c_2(m,n)$. By induction this holds for all non-vanishing coefficients $\tilde{g}^{\star(m,n)}$. 

Thus, below the critical dimension, the flow drives the rescaled potential to a fixed-point potential 
$u_\tau \to u^\star$, which can be represented in the form 
\begin{equation}
	\label{eq:fixedpointseries}
	u^\star(\tchi,\chi) = \sum_{m \ge 1, n \ge 2, m \le n} \frac{1}{m! n!}  \tilde{g}^{\star(m,n)} \tchi^m \chi^n \,.
\end{equation}

In contrast, above the critical dimension,  $u_\tau$ tends to zero. In this case, we  consider  the dimensionful potential $U_{\kappa}$ instead (see Section~\ref{sec:abovedc}). The critical dimension $d_c = 2$, where both potentials, $u_\tau$ and $U_{\kappa}$ tend to zero along the flow, is treated separately at the end of this section.

\subsection{The One-Dimensional Case}

\label{sec:studinone}

Simple scaling arguments (see e.g.~\cite{Hinrichsen:2000p15389}) 
already indicate that the density will behave as
\begin{equation}
 	\label{eq:oneddecay}
	\rho \sim \mathcal{A} t^{-\frac{1}{2}} 
\end{equation}
in the long-time limit, when the dimension $d = 1$, for some amplitude $\mathcal{A}$: The density $\rho$ corresponds to the field $\psi$, such that under renormalization it scales as $\rho = \kappa \tilde\rho$, with the ``dimensionless'' density $\tilde \rho$, see~Eq.~(\ref{eq:rescaledfields}), whereas time scales as $t = \kappa^{-2} \tilde t$, see~Eq.~(\ref{eq:rescaledcoord}). In the following, the most difficult task is to estimate the amplitude $\mathcal{A}$. 

We define the rescaled non-equilibrium force by  $F_\kappa(\psi) :=  \partial_{\tpsi} U_{\kappa}(\tpsi,\psi)|_{\tpsi = 0}$  and its dimensionless counterpart by 
$f_\tau(\chi) := \partial_{\tchi} u_{\tau}(\tchi,\chi)|_{\tchi = 0}$.
Just as the rescaled potential $u_\tau$ flows to $u^\star$, the  renormalization group flow drives $f_\tau$ to its fixed-point value $f^\star$, which according to Eq.~(\ref{eq:fixedpointseries}) may be written as  
\begin{equation*}
	f^\star(\chi) =  \sum_{n \ge 2} \frac{1}{n!} \tilde{g}^{\star(1,n)} \chi^n \,.
\end{equation*}
The kinetic equation becomes 
\begin{equation}
	\partial_t \rho =  - \lim_{\kappa\to0} \kappa^{3} f_\tau(\kappa^{-1} \rho) = - \lim_{\kappa\to0} \kappa^{3} f^\star(\kappa^{-1} \rho)\,,
\label{dtpsi}
\end{equation}
where the second equality is valid to lowest order in $k$. 
The limit must not depend on $\kappa$, since, once the reciprocal scale $\kappa^{-1}$ is much larger than the correlation length, the right hand side of the equation should have converged well. Hence, at the fixed point we will have $f^\star(\chi) \sim c \chi^{3}$, 
when $\chi$ is large, for some universal factor $c$. This implies that the non-equilibrium force $F(\rho) \sim c \rho^3$ and that the kinetic equation~(\ref{eq:genkin}) becomes
\begin{equation*}
	\partial_t \rho = - c \rho^{3} \,, 
\end{equation*}
such that we indeed recover the decay law, Eq.~(\ref{eq:oneddecay}), with $\mathcal{A} = (2 c)^{-\frac{1}{2}}$. 

Determining the factor $c$ 
is tantamount to calculating $f^\star(\chi)$ for large values of $\chi$. This  in turn affords a good knowledge of the fixed-point potential $u^\star(\tchi,\chi)$. 
Typically, the goal of the numerical calculations is to extract critical exponents by considering the flow in the region around the fixed point. In this case, to obtain a satisfactory result, it is often sufficient to perform a series expansion of the Wetterich equation to the first few orders in $\tchi$ and $\chi$ and then to consider the flow of the coefficients $\tilde{g}_\tau^{(m,n)}$. For our problem this clearly will not suffice, since the lower order coefficients only describe the behavior of the force $f^\star$ around the origin but not for large $\chi$. 

\begin{figure}
      \centering \includegraphics[width=0.47\textwidth]{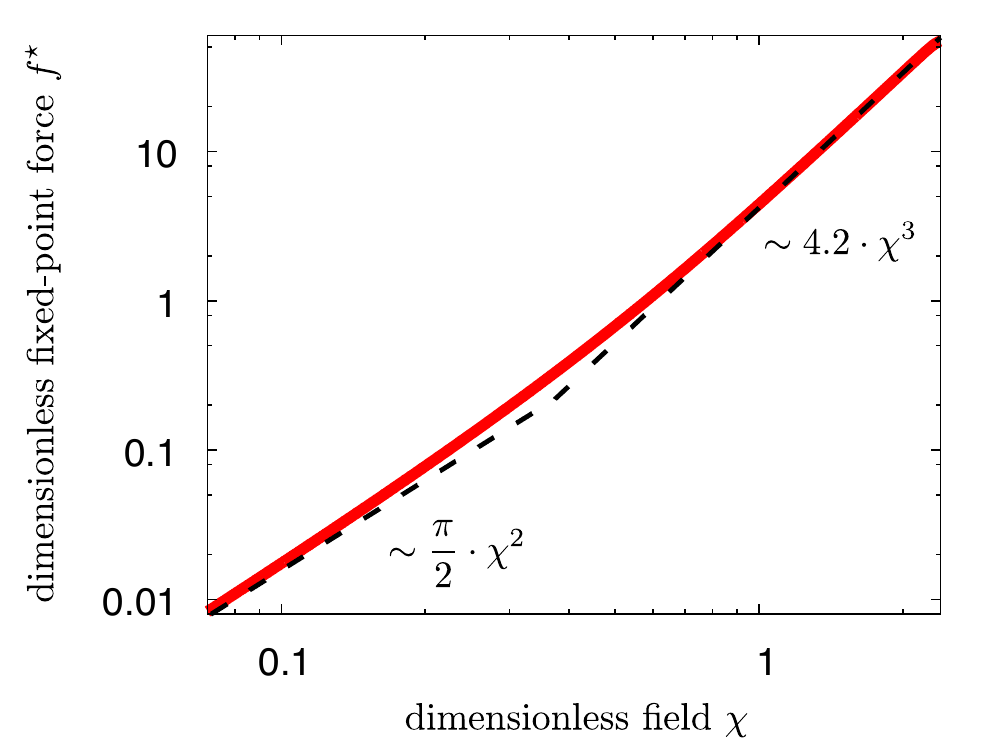}
\caption{Double logarithmic plot of the rescaled non-equilibrium force $f_\tau = f^\star$ (solid red line) at the fixed point for a one-dimensional system. 
 It was obtained by calculating its expansion up to order 125 in $\chi$ within the local potential approximation. For small $\chi$, $f^\star(\chi)  = \frac{\pi}{2} \chi^2 + \frac{\pi^2}{6} \chi^3 + \ldots$ is dominated by the power law $\frac{\pi}{2} \chi^2$ (flat dashed line). When $\chi$ is increased, one reaches a regime where $f^\star$ is described well by $4.2\, \chi^3$  (steep dashed line), before the Pad\'e approximation---utilized to extend the regime of convergence---breaks down. As expected, cf.~Eq.~(\ref{dtpsi}), for large enough $\chi$, in addition to the third order term $\frac{\pi^2}{6} \cdot \chi^3$, the rest of the expansion of $f^\star$ combines to another term scaling as $\chi^3$. }
\label{fig:onedim_powerlaw}
\end{figure}

We have exploited the special simplifications in the flow for the coagulations process to calculate a large number of fixed-point coefficients $\tilde{g}^{\star(m,n)}$.
The equations were solved exactly (yet of course within the truncation of Eq.~(\ref{local_potential_approx})) employing computer algebra software. 
We were thus able to extract the first 125 coefficients $\tilde{g}^{\star(1,n)}$ in the power series of $f^\star$. 
The behavior of $f^\star(\chi)$ for large $\chi$ was evaluated in a double logarithmic plot, cf.~Fig.~\ref{fig:onedim_powerlaw}. 
Since the  power series has a finite radius of convergence we enhanced the result by employing Pad\'e extrapolation~\cite{numericalrecipes}.  
For large values of $\chi$, the terms in the expansion indeed add up a to a power law of the order $\chi^3$. We find that approximately
\begin{equation*}
	f(\chi) \sim 4.2 \, \chi^3 \quad \Rightarrow \quad \rho(t) \sim 0.35\, t^{-\frac{1}{2}}\,.
\end{equation*}
As compared to  the perturbative result
$\mathcal{A} = \frac{1}{2 \pi \epsilon} + \frac{2 \ln(8 \pi) - 5}{8 \pi}\approx 0.22$ of~\cite{Lee:1994p10274} (with $\epsilon = d_c - d = 1$), this is much closer to the exact decay amplitude $\frac{1}{\sqrt{ 2 \pi}} \approx 0.40$~\cite{Bramson:1980p11383,Torney:1983p15519,Lushnikov:1987p22899,Spouge:1988p23035,Doering:1988p15522,BenAvraham:1998p23566}.

\subsection{Generalization to $d< d_c$}
\label{subsec:belowdc}

\begin{figure}
      \centering \includegraphics[width=0.5\textwidth]{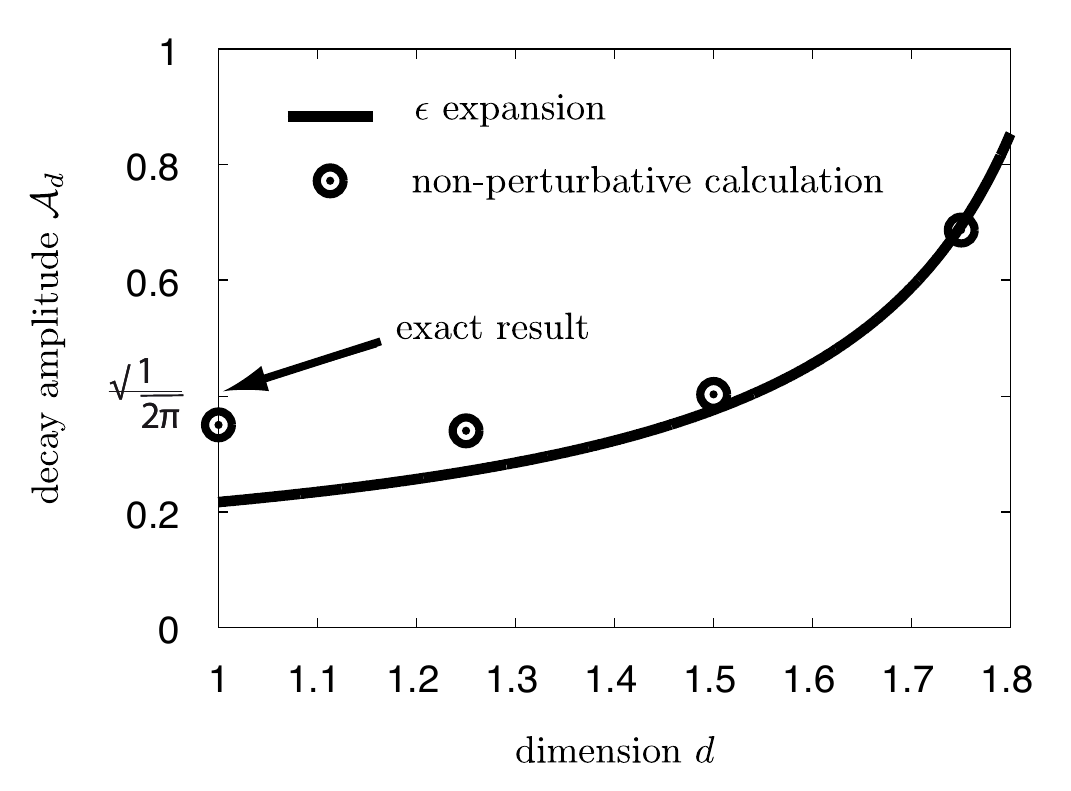}
\caption{Results for the amplitude $\mathcal{A}_d$ of the long-time decay $\rho \sim \mathcal{A}_d \rho^{-\frac{d}{2}}$ ($d<2$). The circles are our estimates from the non-perturbative renormalization group calculations within the local potential approximation. They are compared to the exact result $\mathcal{A}_1 = \frac{1}{\sqrt{2 \pi}}$ in one dimension~\cite{Bramson:1980p11383} and to the perturbative result $\frac{1}{2 \pi \epsilon} + \frac{2 \ln(8 \pi) - 5}{8 \pi}$ of~\cite{Lee:1994p10274} (solid line). The latter is an expansion in the deviation $\epsilon$ from the upper critical dimension, $\epsilon = d_c - d$. It is doubtful whether $\epsilon = 1$ (corresponding to one dimension) can be considered small. Indeed, despite the relatively crude truncation,  the non-perturbative approach provides a substantially better result.}
\label{fig:d_below_dc}
\end{figure}

Formally one can extend the above approach to ``dimensions'' $d$ below the critical dimension $d_c = 2$ and calculate the corresponding amplitude $\mathcal{A}_d$ in the long-time scaling of the density $\rho \sim \mathcal{A}_d \cdot t^{-\frac{d}{2}}$ (see Figure \ref{fig:d_below_dc}). In complete analogy to the previous section, we find that for large values of the field $\chi$, the fixed-point result of the non-equilibrium force must scale as 
	$f^\star(\chi) \sim c_d \chi^{\frac{d+2}{d}}$ 
with the dimension-dependent but otherwise universal factor $c_d$. Hence for the kinetic equation we have
\begin{equation*}
	\partial_t \rho = - F(\rho) \sim - c_d \rho^{\frac{d+2}{d}} \,, \quad  \rho \sim \left(\frac{2 c_d t}{d}\right)^{-\frac{d}{2}} \,. 
\end{equation*}

As for the one-dimensional case, the factor $c_d$ is gleaned from a sufficient number of coefficients in the expansion of the rescaled non-equilibrium force $f_\tau(\tchi,\chi) = f^\star(\tchi,\chi)$ at the fixed point. 
One observes that the Pad\'e approximation works the better, i.e.~converges for larger values of $\chi$, the closer one approaches the critical dimension. Indeed, from perturbation theory one expects that near the critical dimension only $\tilde{g}^{\star (2,2)}$ and $\tilde{g}^{\star (1,2)}$ are important, so that the approximation should converge quickly.
Performing the limit $d_c -d = \epsilon \to 0$, we can make contact with a result from perturbation theory~\cite{Lee:1994p10274}. To this end, we assume that to lowest order  
\begin{equation*}
	f^\star(\chi) \sim c_d \chi^{\frac{4-\epsilon}{2-\epsilon}} \sim \left(\tilde{\lambda}^\star + O(\epsilon^2)\right) \chi^{\frac{4-\epsilon}{2-\epsilon}} \,,
\end{equation*}	
i.e.~the constant $c_d$ is, to good approximation, equal to the coupling $\tilde{\lambda}^\star = \frac{1}{2} \tilde{g}^{\star (1,2)}$. This is plausible since the exponent  $\frac{4-\epsilon}{2-\epsilon} \approx 2$. 
Our assumption indeed allows to derive from Eq.~(\ref{lambda_fixed}) the relation $c_d = 2 \pi \epsilon + O(\epsilon^2)$, and thus we recover the result from perturbation theory $\mathcal{A}_d = \frac{1}{2 \pi \epsilon}$ to leading order in $\epsilon$.

\subsection{Treatment for the Critical Dimension}

At the critical dimension $d_c = 2$ the couplings $\tilde{g}_\tau^{(1,2)}, \tilde{g}_\tau^{(2,2)}$ behave as $\frac{1}{\tau}$ when $\tau \to - \infty$. Since the Feynman diagrams which determine the flow of $\tilde{g}_\tau^{(m,n)}$ involve $n$ of these ``elementary'' couplings, the other coefficients go to zero as $\frac{1}{\tau^n}$. Therefore, the potential vanishes along the renormalization group flow, $u_\tau \to 0$, we cannot take the limit of Eq.~(\ref{dtpsi}) and a straightforward application of the analysis of the previous sections to determine the long-time behavior of the density is not possible.  
Instead, we start, at finite renormalization time $\tau$, with the dimensionless equation 
\begin{equation}
\label{eq:crtitdimeqmot}
	\partial_{\tilde t} \chi = - \left[\chi +f _\tau(\chi)\right]  \quad(\tilde t = \kappa^2 t) \,,
\end{equation}
where the constant term $\sim \chi$ stems from the dimensionless cutoff function (\ref{eq:dimlesslitim}) at vanishing rescaled momentum $\tilde{q} = 0$ ($\lim_{\tilde{q} \to 0} \tilde{q}^2 r(\tilde{q}^2) \to 1$). In this equation the long-range fluctuations have not yet been integrated out completely. At finite $\tau$ the results roughly correspond to those of a system of finite size, with edge length $\kappa^{-1}$ (recall that $\kappa = \Lambda \exp(\tau)$).  

In the previous subsection we saw that, as the critical dimension is approached, $d \to 2$, the decay amplitude $\mathcal{A}_d$ is determined by the lowest order coefficient $\tilde{\lambda}_\tau = \frac{1}{2} \tilde{g}^{(1,2)}_\tau$ (up to corrections in the difference $\epsilon = 2 - d$). Therefore, let us assume that we may set 
\begin{equation*}
	f _\tau(\chi) = \tilde{\lambda}_\tau \chi^2 = - \frac{2 \pi}{\tau} \chi^2 \,,
\end{equation*}
ignoring higher order coefficients (for a more rigorous analysis by means of the perturbative renormalization group, we refer to~\cite{Lee:1994p10274}). 
A constant initial density $\rho_0$ implies a diverging dimensionless initial density $\frac{\rho_0}{\kappa^2}$. Solving   Eq.~(\ref{eq:crtitdimeqmot}) for this initial condition, we obtain
\begin{equation*}
	\chi(\tilde{t}) = \frac{\rho_0 \tau}{- 2 \pi \tilde{t} \rho_0 + \tau \kappa^2} \,,
\end{equation*}
as long as $\chi$ is large enough so that the term $-\chi$ in Eq.~(\ref{eq:crtitdimeqmot}) can be disregarded. (It destroys the algebraic behavior for large times, where $\chi(\tilde{t}) \sim \exp(- \tilde{t})$ decays exponentially, as one would expect for a finite-size system.)
Thus, after an initial transient time of the order $\tau \kappa^2$, which goes to zero exponentially fast as $\tau \to -\infty$ and can therefore be neglected, we have $\chi(\tilde{t}) = - \frac{\tau}{2 \pi \tilde{t}}$.  Inserting $\chi(\tilde t) = \frac{\rho(t)}{\kappa^2}$ and $\tilde{t} = \kappa^2 t$ (see Eqs.~(\ref{eq:rescaledcoord},\ref{eq:rescaledfields}) with $\psi = \rho$) and $\tau = \ln(\kappa/\Lambda) \sim - \frac{\ln(t)}{2}$, we recover the result~\cite{Lee:1994p10274}
\begin{equation*}
	\rho(t) = \frac{\ln t}{4 \pi t} \,,
\end{equation*}
for large times $t$.

\section{Behavior Above the Critical Dimension}

\label{sec:abovedc}

We have seen 
that below the  critical dimension two the renormalization group flow drives the (dimensionful) renormalized decay rate $\lambda_\kappa := \frac{1}{2} g_\kappa^{(1,2)}$ to zero. In order to treat this singularity, we introduced the dimensionless couplings $\tilde{g}_\kappa^{(m,n)} = \kappa^{-2+d(n-1)} g_\kappa^{(m,n)}$, which tend to a finite, universal value $\tilde{g}^{\star(n,m)}$ as the infrared cutoff scale $\kappa$ becomes small. Without tuning of parameters, the couplings flow to a fixed point, which implies anomalous long-time behavior $\rho \sim \mathcal{A}_d t^{-\frac{d}{2}}$ with a universal amplitude $\mathcal{A}_d$. We shall show in the following that in contrast to this, above the critical dimension the renormalized decay rate $\lambda_\kappa$ attains a finite value as the scale $\kappa$ goes to zero. 
Thus, above the critical dimension the long-time behavior obeys the LMA $\partial_t \rho \sim  - \mu \rho^2$ with a non-universal macroscopic decay rate $\mu := \lambda_{\kappa=0}$ 
and at long times we recover  the ``classical'' scaling $\rho \sim \mu^{-1} t^{-1}$. 

\subsection{Derivation of the Macroscopic Decay Rate}

The renormalized reaction rate can be obtained from the identity 
\begin{equation*}
	\lambda_\kappa = \frac{1}{2} \Gamma^{(1,2)}_{\kappa\,(\mathbf{0},0;\mathbf{0},0;\mathbf{0},0)} (2\pi)^{3d+3} (V T)^{-1} \,, 
\end{equation*}
where $V = \sum_{\mathbf{x}} = (2\pi)^d \delta(\mathbf{0})$ and $T = \int \! \mathrm{d}t = 2\pi \delta(0)$ denote the asymptotically large volumes in space and time which are summed and integrated over, respectively (in the final result they drop out).  

\begin{figure}
      \centering \includegraphics[width=0.5\textwidth]{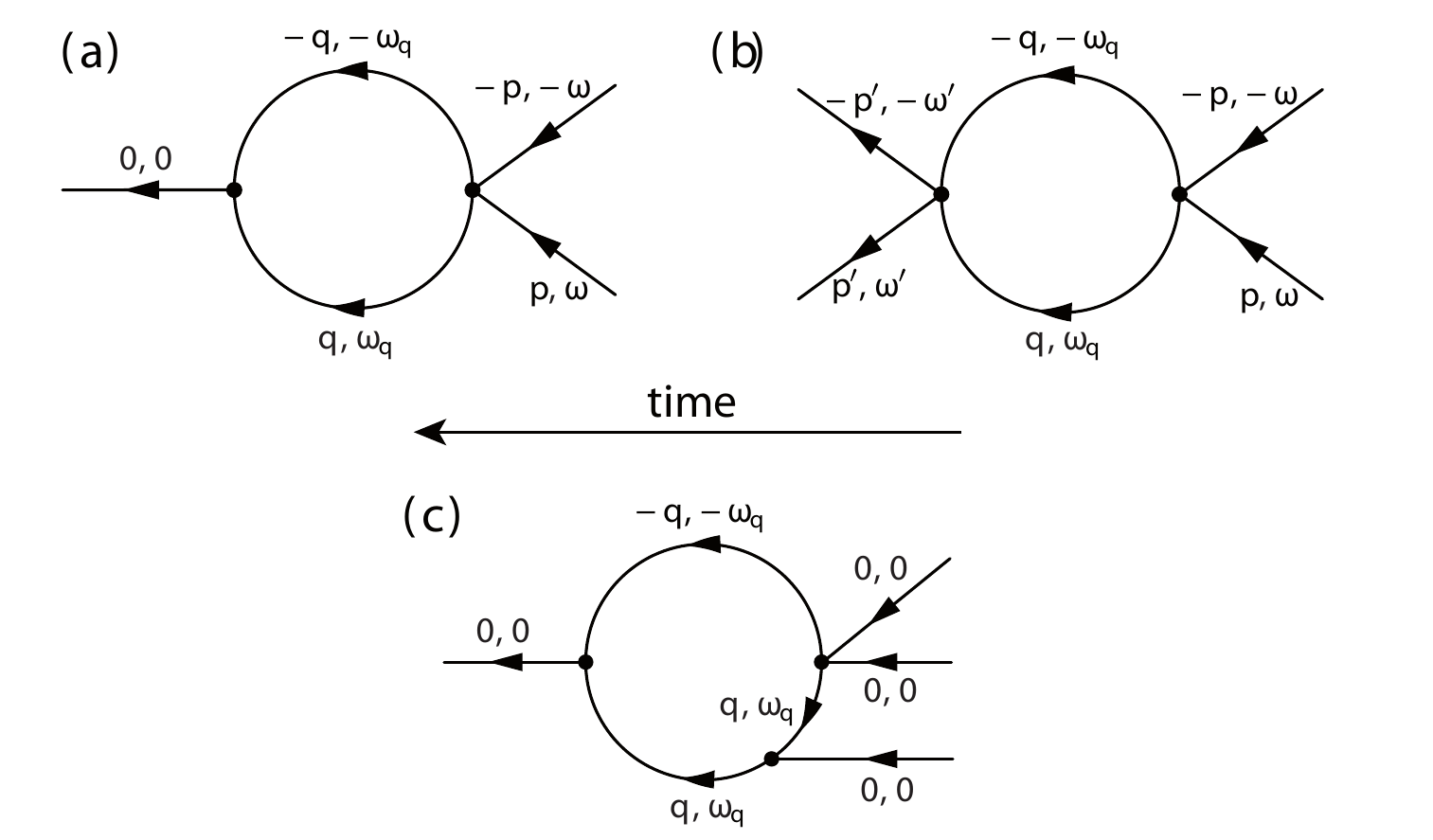}
\caption{In (a) and (b) the one-loop Feynman diagrams for the calculation of the flow of the decay rate $\lambda_k = \frac{1}{2} g_k^{(1,2)} = \frac{1}{2} \Gamma^{(1,2)}_{\kappa\,(\mathbf{0},0;\mathbf{0},0;\mathbf{0},0)} (2\pi)^{12} (V T)^{-1}$ are shown. Diagram $(a)$ determines the flow of the (1,2)-vertex function $\Gamma^{(1,2)}_{\kappa\,(\mathbf{0},0;\mathbf{p},\omega;-\mathbf{p},-\omega)}$ and involves a (2,2)-vertex function $\Gamma^{(2,2)}_{\kappa\,(\mathbf{p}^\prime,\omega^\prime,-\mathbf{p}^\prime,-\omega^\prime;\mathbf{p},\omega;-\mathbf{p},-\omega)}$, whose flow (deduced from diagram $(b)$) provides a  closed and exact formula for the (2,2)-vertex function. Diagram $(c)$ determines the flow of the coefficient $g_k^{(1,3)}$. The three internal propagators imply a factor $\left(\frac{1}{k^2}\right)^3$ for the behavior of $g_k^{(1,3)}$ in the limit of small $k$.  Integration over momentum and frequency space in the Wetterich equation abates this singularity by a factor $k^{d+2}$. Hence, when $2 < d < 4$ the coefficient $g_k^{(1,3)}$ diverges as $k^{d-4}$. Above four dimensions it approaches a finite value as $k \to 0$.
}
\label{fig:feyn2}
\end{figure}

Let us first restrict to the case where there is only local interaction between the particles, $\lambda(\by - \bx) = \delta_{\bx,\by} \lambda$, so that they can be regarded as extending over one site of the lattice. To calculate $\Gamma^{(1,2)}_{\kappa\,(\mathbf{0},0;\mathbf{0},0;\mathbf{0},0)}$ we consider  the flow of the more general vertex function $\Gamma^{(1,2)}_{\kappa\,(\mathbf{0},0;\mathbf{-p},-\omega;\mathbf{p},\omega)}$. It is determined by the one-loop diagram~\ref{fig:feyn2}$(a)$ and apart from a (1,2)-vertex contains also a (2,2)-vertex. The flow of the  (2,2)-vertex function $\Gamma^{(2,2)}_{\kappa\,(\mathbf{p}^\prime,\omega^\prime;-\mathbf{p}^\prime,-\omega^\prime;\mathbf{p},\omega;-\mathbf{p},-\omega)}$ follows from diagram~\ref{fig:feyn2}$(b)$. It is self-contained in the sense that it only depends on the (2,2)-vertex function itself. Since the internal momenta and frequencies of the one-loop diagram do not depend on the external ones, we have that the identity 
$\Gamma^{(2,2)}_{\kappa\,(\mathbf{p}^\prime,\omega^\prime;-\mathbf{p}^\prime,-\omega^\prime;\mathbf{p},\omega;-\mathbf{p},-\omega)} = \Gamma^{(2,2)}_{\kappa\,(\mathbf{0},0;\mathbf{0},0;\mathbf{0},0;\mathbf{0},0)}$, which holds at $\kappa = \infty$, is bequeathed to all scales $\kappa$. By the same token, the identity 
$\Gamma^{(1,2)}_ {\kappa\,(\mathbf{0},0;\mathbf{p},\omega;-\mathbf{p},-\omega)} = \Gamma^{(2,2)}_ {\kappa\,(\mathbf{0},0;\mathbf{0},0;\mathbf{p},\omega;-\mathbf{p},-\omega)} \frac{(2 \pi)^{d+1}}{2}$, which evidently holds for the microscopic action~(\ref{eq:actionannicoag}), remains valid along the renormalization group flow.
Thus, evaluating diagram~\ref{fig:feyn2}$(b)$, we obtain the flow for the renormalized decay rate 
\begin{equation*}
	\partial_\kappa \lambda_\kappa =  
	- 2 \tilde\partial_\kappa \int_{\mathbf{q},\omega_{q}} \frac{\lambda_\kappa^2}{\left(R_\kappa(\mathbf{q}) + \epsilon(\mathbf{q})\right)^2 + \omega_{q}^2}  = 
\end{equation*}
\begin{equation*}
	= 2 \lambda_\kappa^2 \frac{\int_\mathbf{q} \Theta(\kappa^2 - \epsilon(\mathbf{q}))}{\kappa^3}\,.
\end{equation*}
Therefore, in the case of one-site objects the exact solution to the macroscopic decay rate reads
\begin{equation*}
\label{eq:onesite}
	\frac{1}{\mu}  = \frac{1}{\lambda_0} = \frac{1}{\lambda} +  \int_{\mathbf{q}} \frac{1}{\epsilon(\mathbf{q})} \,.
\end{equation*}
This equation connects the  microscopic reaction rate $\lambda$ with its macroscopic counterpart $\mu$ by a term which depends on the structure of the lattice. 
The solution is finite on condition that the dimension $d>2$. In contrast, when $d\le 2$ the integral diverges to infinity, indicating an anomalously slow decay, which was discussed in the previous section. For a cubic lattice with lattice spacing $a=1$ and diffusion constant $D=1$, by numerical integration we find that $\mu^{-1} = \lambda^{-1} + 0.252731009858(3)$. This value is corroborated by our numerical simulations, where we have considered the long-time decay of the density $\rho \sim \mu^{-1} t^{-1}$~\cite{Winkler:2012p23673}.

We now proceed to derive the flow equation to the renormalized reaction rate $\lambda_\kappa$ 
for general reaction kernels $\lambda(\bz)$, whose interaction may extend over several sites. 
Similar as for one-site objects, we have the identity $\Gamma^{(1,2)}_ {\kappa\,(\mathbf{0},0;\mathbf{p},\omega;-\mathbf{p},-\omega)} = \Gamma^{(2,2)}_ {\kappa\,(\mathbf{0},0;\mathbf{0},0;\mathbf{p},\omega;-\mathbf{p},-\omega)} \frac{(2 \pi)^{d+1}}{2}$. 
Thus, again it suffices to calculate the (2,2)-vertex function $\Gamma^{(2,2)}_{\kappa\,(\mathbf{p}^\prime,\omega^\prime;-\mathbf{p}^\prime,-\omega^\prime;\mathbf{p},\omega;-\mathbf{p},-\omega)}$,  with the important difference, however, that in general this vertex function depends explicitly on both of the momenta $\mathbf{p}^\prime$ and $\mathbf{p}$,  
\begin{equation*}
	\Gamma^{(2,2)}_{\kappa\,(\mathbf{p^\prime},\omega^\prime;-\mathbf{p^\prime},-\omega^\prime;\mathbf{p},\omega;-\mathbf{p},-\omega)} 
	\equiv \frac{4 V T}{(2 \pi)^{4 d+4}} \lambda(\mathbf{p}^\prime,\mathbf{p}) \,.
\end{equation*}
Notice that there is no dependence on the frequencies $\omega^\prime$ and $\omega$. We obtain 
\begin{equation*}
	\partial_\kappa \lambda_\kappa(\mathbf{p}^\prime,\mathbf{p}) = 
	2 \int_{\mathbf{q}} \lambda_\kappa(\mathbf{p}^\prime, \mathbf{q}) \lambda_\kappa (\mathbf{q},\mathbf{p}) \frac{\Theta(\kappa^2 - \epsilon(\mathbf{q}))}{\kappa^3} \,.
	\label{eq:extp}
\end{equation*}
Equivalently, in position space 
\begin{equation}
	\partial_\kappa \lambda_\kappa(\mathbf{x},\mathbf{y}) = 
	2 \int_{\mathbf{q}} \lambda_\kappa(\mathbf{x}, \mathbf{q}) \lambda_\kappa (\mathbf{q},\mathbf{y}) \frac{\Theta(\kappa^2 - \epsilon(\mathbf{q}))}{\kappa^3} 
	\,,
	\label{eq:ext}
\end{equation}
which may be rewritten as
\begin{IEEEeqnarray*}{rCl}
	\partial_\kappa \lambda_\kappa(\mathbf{x},\mathbf{y}) &=& 
	\frac{2}{k^3} \sum_{\mathbf{z}}  \lambda_\kappa(\mathbf{x},\mathbf{z}) \left( \mathcal{P}_1 \circ \lambda_\kappa\right)\!(\mathbf{z},\mathbf{y}) \label{eq:ext1}
	 \\
	&=& 
	\frac{2}{k^3} \sum_{\mathbf{z}}  \left( \mathcal{P}_2 \circ \lambda_\kappa\right)\!(\mathbf{x},\mathbf{z}) \lambda_\kappa(\mathbf{z},\mathbf{y}) 
	\,, \label{eq:ext2}
\end{IEEEeqnarray*}
where the projections $\mathcal{P}_1$ and $\mathcal{P}_2$ are defined by  
\begin{equation*}
	\left(\mathcal{P}_1 \circ \lambda_\kappa\right)(\mathbf{z},\mathbf{y}) = \int_{\mathbf{q}} \exp( i \mathbf{q} \cdot \mathbf{z})  \lambda_\kappa(\mathbf{q},\mathbf{y}) \Theta(\kappa^2 - \epsilon(\mathbf{q})) \,,
\end{equation*}
\begin{equation*}
	\left(\mathcal{P}_2 \circ \lambda_\kappa\right)(\mathbf{x},\mathbf{z}) = \int_{\mathbf{q}} \exp( i \mathbf{q} \cdot \mathbf{z})  \lambda_\kappa(\mathbf{x},\mathbf{q}) \Theta(\kappa^2 - \epsilon(\mathbf{q})) \,.
\end{equation*}
For numerical calculations it is an important simplification that the support of $\lambda_\kappa(\mathbf{x},\mathbf{y})$ is contained in $S \times S$, where $S$ is the support of $\lambda(\mathbf{x})$. 

 As an example for extended objects, consider balls of radius $\frac{R}{2}$ in three-dimensional continuum space. The reaction kernel $\lambda(\bz) = \lambda \Theta(R-z)$,  where $z$ is the distance between the centers of the particles. Despite the conservation of the support, Eq.~(\ref{eq:ext}) is still difficult to solve for general $\lambda$. However, a particularly interesting case are instantaneous reactions, obtained by taking the limit $\lambda \to \infty$.  This corresponds to the classical problem treated by Smoluchowski~\cite{Smoluchowski:1917p16756}, who argued that $\mu = 4 \pi R D$ (where $D$ denotes the diffusion constant), by a heuristic approach, which was later rigorously confirmed by Doi~\cite{Doi:1976p15932}. Notice that the problem is equivalent to $\lambda(\bz) = \lambda \, \delta(R-z)$, where the kernel  is nonzero only on the surface of a sphere, in the limit of infinitely fast reaction, $\lambda \to \infty$. We now derive a simplified exact flow equation for this and similar cases. 

Suppose that the kernel $\lambda_\kappa(\mathbf{x})=\sum_{\mathbf{y}} \lambda_\kappa(\mathbf{x},\mathbf{y}) = \sum_{\mathbf{y}} \lambda_\kappa(\mathbf{y},\mathbf{x})$ can be described by only one degree of freedom, i.e.~on its support $S$, the reaction kernel $\lambda_\kappa(\mathbf{x})$ is a constant (for fixed scale $\kappa$) and all the elements of the support are equivalent, in the sense that every element of the support can be mapped onto every other element under rotations and reflections that conserve the action~$\mathcal{S}$. 

According to Eq.~(\ref{eq:ext1})
\begin{equation*}
	\partial_\kappa \lambda_\kappa(\mathbf{x}) = \frac{2}{k^3} \sum_{\mathbf{z}} \lambda_\kappa(\mathbf{x},\mathbf{z}) \sum_{\mathbf{y}}(\mathcal{P}_1 \circ \lambda_\kappa)(\mathbf{z},\mathbf{y}) = 
\end{equation*}
\begin{equation}
	= \frac{2}{k^3} \sum_{\mathbf{z} \in S} \lambda_\kappa (\mathbf{x},\mathbf{z}) (\mathcal{P} \circ \lambda_\kappa)(\mathbf{z}) \,,
	\label{eq:simple}
\end{equation}
with the projection 
\begin{equation}
\label{eq:projection}
\left(\mathcal{P}\circ\lambda_\kappa\right)\!(\mathbf{z}) = \int_{\mathbf{q}} \exp( i \mathbf{q} \cdot \mathbf{z})  \lambda_\kappa(\mathbf{q}) \Theta(\kappa^2 - \epsilon(\mathbf{q})) \,.
\end{equation}
Since $(\mathcal{P} \circ \lambda_\kappa)(\mathbf{z})$ is a constant on $S$, the sum in Eq.~(\ref{eq:simple}) is trivial and we obtain
\begin{equation}
	\label{eq:flow_lambda_x}
	\partial_\kappa \lambda_\kappa(\mathbf{x}) = \frac{2 \lambda_\kappa(\mathbf{x}) \left(\mathcal{P}\circ\lambda_\kappa\right)\!(\mathbf{x})}{\kappa^3}\,.
\end{equation}
Alternatively this can be written as 
\begin{equation}
	\partial_\kappa \lambda_\kappa(\mathbf{p}) = 2 \int_{\mathbf{q}} \, \lambda_\kappa(\mathbf{p}-\mathbf{q}) \lambda_\kappa(\mathbf{q}) \frac{\Theta(\kappa^2 - \epsilon(\mathbf{q}))}{\kappa^3}\,.
	\label{eq:flow_lambda_p}
\end{equation}

As elaborated in the Appendix, from Eq.~(\ref{eq:flow_lambda_x}) one obtains an analytic solution for the kernel $\lambda(\bz) = \lambda \, \delta(R-z)$ in continuum space. In particular, we recover Smoluchowski's classical result $\mu = 4 \pi D R$ for the macroscopic decay rate of spheres of radius $R$, diffusing with diffusion constant $D$ and coagulating upon contact.  In the stochastic simulations we found it  preferable to work with objects defined on a lattice rather than in continuous space. In Fig.~\ref{fig:extended_objects} we give two examples of extended object defined on a lattice,  whose exact macroscopic decay rate can be derived from Eq.~(\ref{eq:flow_lambda_p}).

\subsection{Universal Correction to the Law of Mass Action}

\label{sec:unicorr}

To study corrections to the LMA, let us represent the non-equilibrium force as the limit of a power series 
\begin{equation*}
	F(\rho) = \lim_{\kappa\to 0}  \sum_{n\ge 2} \frac{1}{n!} g_\kappa^{(1,n)} \rho^n \,,
\end{equation*}
exploiting the fact that $\Gamma_\kappa$ is analytic if $\kappa>0$~\cite{Berges:2002p16426}. 
The lowest order coefficient $g_\kappa^{(1,2)} = 2 \lambda_\kappa$, treated above, converges to a constant value, the macroscopic decay rate $g_{\kappa=0}^{(1,2)} = 2 \mu$. Thus, assuming that the higher order terms $\frac{1}{n!} g_\kappa^{(1,n)} \rho^n$ ($n > 2$) can be neglected, we would recover the LMA term $F(\rho) = \mu \rho^2$, quadratic in the density $\rho$. 
It is crucial to notice that the next to leading terms are not  simply $\frac{1}{3!} g_{\kappa=0}^{(1,3)} \rho^3 + \frac{1}{4!} g_{\kappa=0}^{(1,4)} \rho^4 + \ldots$
Rather, in $d$ dimensions \emph{all} coefficients $g_{\kappa}^{(1,n)}$, with $n \ge \frac{d + 2}{2}$, turn out to diverge as $\kappa$ goes to zero.
In the following analysis, we show that the infinite sum of these diverging terms converges and gives the finite contribution  $\sum_{n\ge \frac{d+2}{2}} \frac{1}{n!} g_\kappa^{(1,n)} \rho^n \sim c_d \rho^{\frac{d+2}{2}}$ (up to possible logarithmic factors in $\rho$), for some constant $c_d$. In three dimensions this provides a relatively large leading correction.   

The flow of the physically relevant couplings $g_{\kappa}^{(1,n)}$ 
 is calculated from diagrams with $n$ incoming and one outgoing leg.  
To lowest order, their divergence stems from the contribution to the flow of diagrams which only contain (1,2)- and (2,2)-vertices and  follows from power counting. Let us exemplify this for the (1,3)-coupling $g_\kappa^{(1,3)}$, with the associated one-loop diagram of Fig.~\ref{fig:feyn2}(c). 
For any finite scale $\kappa$, also the couplings must be finite. Therefore, the divergence in $\kappa$ of $g_\kappa^{(1,3)}$ builds up in the limit of small $\kappa$, where only long-wavelength and short-frequency fluctuations,  $q \lesssim \kappa$, and $\omega \lesssim \kappa^2$, contribute to the flow. 
Above the critical dimension, the (1,2)- and (2,2)-vertex functions are not divergent for $\kappa= q=\omega=0$, but attain a finite value, which is equal to the macroscopic decay rate $\mu$. Hence we can take their value at  $\kappa = q = \omega = 0$, if we are only interested in  the strongest divergence.
The evaluation of the diagram then yields
\begin{equation}
	\partial_{\kappa} g_{\kappa}^{(1,3)} = - \tilde\partial_\kappa 16 \int_{\bq,\omega} G_\kappa(\bq,\omega)^2 G_\kappa(-\bq,-\omega) (- \mu)^3 \,
	\label{eq:logdiv}
\end{equation}	
with the propagator
\begin{equation*}
	 G_\kappa(\bq,\omega) = \frac{1}{\left(\kappa^2 - \epsilon(\bq)\right) \Theta\left(\kappa^2 - \epsilon(\bq)\right) + \epsilon(\bq) + i \omega} \,,
\end{equation*}
and the macroscopic reaction decay rate $\mu = \mu(\bq = \mathbf{0})$. 
Due to the derivative $\tilde \partial_\kappa = \partial_\kappa R_\kappa \cdot \partial_{R_\kappa}$, the integration is restricted to the domain $\epsilon(\bq) < \kappa^2$, where the propagator is independent of $\bq$, i.e.~
\begin{equation*}
	\tilde\partial_{\kappa} \int_\bq = \tilde \partial_{\kappa} \int_{\epsilon(\bq) < \kappa^2} \,,
\end{equation*}
such that within the domain of integration the propagator simplifies to 
\begin{equation*}
	G_\kappa(\bq,\omega) = \frac{1}{\kappa^2 + i\omega} \,.
	\label{eq:propasimple}
\end{equation*}
The fact that the integral, Eq.~(\ref{eq:logdiv}), does not depend on the full reaction kernel $\mu(\bq)$ already indicates that these divergences cannot depend on the shape and size of the objects:  Originating in long{\strich}wavelength fluctuations around $q = 0$, they do not resolve the details of the reaction kernel.

To lowest order in $\kappa$, the dispersion relation $\epsilon(\bq)$ can be approximated by the ``continuum limit'' $\epsilon(\bq) = q^2$. Therefore, the divergences are not only unaffected by the shape and size of the particles, but also independent of the structure of the lattice; it is as if the divergent terms only ``see'' structureless point particles (for which $\mu(\bq) = \mu$, independent of the momentum $\bq$) that are embedded in continuous space (where the dispersion is simply $\epsilon(\bq) = q^2$).  
To lowest order in $\kappa$ the integration over the momentum $\bq$ then yields the volume of the $d$-dimensional ball with radius $\kappa$.  Thus, up to  some positive constant factor,   Eq.~(\ref{eq:logdiv}) becomes
\begin{equation*} 
	\mu^3 \kappa^d   \int \!\mathrm{d} \nu\,   \frac{\kappa^2}{\left(1 +  i \nu\right)^2 \left(1 - i \nu\right)} \partial_\kappa \left(\frac{1}{\kappa^2}\right)^3 \,.
\end{equation*}	
The integral, where we substituted $\frac{\omega}{\kappa^2} = \nu$, contributes a positive factor. 
Thus, the coupling $g_{\kappa}^{(1,3)}$ scales as $\kappa^{d-4}$ for dimension  $d < 4$, diverges logarithmically at $d=4$ (in both of these cases $g_{\kappa}^{(1,3)}$ diverges to positive infinity), and converges to a finite value for $d>4$.  

In summary, and generalizing to arbitrary couplings, the strongest divergence of a diagram is obtained by the following prescription. Each propagator gives a factor 
\begin{equation*}
	G_{\kappa}(\bq,\omega) \sim \frac{1}{\kappa^2} \,.
	\label{eq:pres1}
\end{equation*}
The resulting divergence is attenuated by the integration over the momenta $\bq$ and frequencies 
\begin{equation}
	\int \! \md^d q \sim \kappa^d \,, \quad \int \! \md \omega \, \sim \kappa^2 \,,
	\label{eq:prescriptint}
\end{equation}
where in  the left hand formula the exponent 3 is simply the dimension. Finally, the lowest order contribution in $\kappa$  of the vertices must be multiplied with the result, in particular (1,2)- and (2,2)-vertices give rise to the constant factor $-\mu$. 

It is now straightforward to determine the strongest divergences of the couplings $g_\kappa^{(1,n)}$. We need to consider the one-loop diagrams with $n$ incoming and one outgoing leg that contain only (1,2)- and (2,2)-vertices. Clearly, in these diagrams there are exactly $n$ of these vertices, connected by $n$ propagators, which gives a factor $(-\mu)^n \kappa^{-2 n}$. After integrating over the momenta and frequencies, cf.~Eq.~(\ref{eq:prescriptint}), we have that the couplings $g_\kappa^{(1,n)}$ converge to a finite value $g_0^{(1,n)}$ if $2n - (d+2) < 0$ and otherwise diverge, 
\begin{equation} 
\label{eq:alternatingseq}
	g_\kappa^{(1,n)} \sim  \left\{
	\begin{array}{cl}
		 g_0^{(1,n)}  & \text{if  }\, 2n  < d+2  \,, \\ 
		 (-1)^{n}  c_{d,n} \mu^n \ln(\kappa) & \text{if  }\, 2n  =  d +2 \,, \\
		 (-1)^{n+1}  c_{d,n}  \mu^n \kappa^{d+2-2n} & \text{if  }\, 2n > d+2  \,,
	\end{array} \right. 
\end{equation}
with some $c_{d,n} > 0$. 

Diagrams with higher order vertices (more than two incoming legs) can be neglected. Suppose, for instance, that such a one-loop diagram contains a $(1,n^\prime)$-vertex ($n^\prime > 2$), which contributes a factor $\kappa^{d+2 - 2n^\prime}$  ($2n^\prime > d + 2$). If we replace this vertex by a string of $n^\prime-1$ (1,2)-vertices, 
connected one by  one by $n^\prime-2$ propagators,  this results in the more relevant factor of the order $\kappa^{-2(n^\prime-2)}$ (recall that the dimension is larger than the critical dimension $d_c = 2$).  

The fact that the couplings $g_\kappa^{(1,n)}$ diverge in an alternating sequence is not surprising, given that one also obtains such sequences when one takes the Taylor expansion of the non-analytic function $x^\alpha$ for a non-integer $\alpha > 0$ around, say, $x=1$. Indeed, we are on the lookout for such non-analytic terms. The infinite sum of diverging terms can be written as  
\begin{equation}
\label{eq:originalinfsum2}
	\sum_{2 n > d + 2} \frac{1}{n!} g_{\kappa}^{(1,n)} \rho^n \sim 
\kappa^{d+2} f_d\!\left(\frac{\mu \rho}{\kappa^2}\right)\,,
\end{equation}
for some scaling function $f_d$. Since for large systems the non-equilibrium force must become independent of the system size, corresponding to the reciprocal scale $\kappa^{-1}$, one obtains
\begin{equation*}
	f_d(x) \sim c_d x^{\frac{d+2}{2}} \,,
\end{equation*}
for some constant $c_d$. In particular, this indicates  that in three dimensions, as opposed to higher dimensions, the next to leading term to the LMA term in the force $F$ is not of the order three but a non-analytic term of order~$\frac{5}{2}$, 
\[
	F(\rho) = \mu \rho^2 + c_3  (\mu\rho)^{\frac{5}{2}} + \ldots
\]

In fact, we cannot strictly rule out that the sum in Eq.~(\ref{eq:originalinfsum2}) does depend on $\kappa$.  The $\kappa$-dependent can cancel with some other correction term (or terms) to the potential. Actually, this can rectify the problem of the logarithmic correction in even dimensions. Assuming that 
\begin{equation*}
	f_{\kappa}\left(x\right) \sim c_d \cdot x^{\frac{d}{2}+1} \ln(x)\,, 
\end{equation*}
for even dimensions $d$, then the sum in Eq.~(\ref{eq:originalinfsum2}), in addition to a $\kappa$-independent term $c_d (\mu \psi)^{\frac{d}{2}+1} \ln(\mu \psi)$, gives rise to a term $- 2 c_d (\mu \psi)^{\frac{d}{2}+1} \ln(\kappa)$. This term can cancel with the term that is logarithmic in $\kappa$, cf.~Eq.~(\ref{eq:alternatingseq}).

Let us finally exploit our findings to calculate the correction term \emph{exactly} from the Wetterich equation. 
We choose the ansatz
\begin{equation}
\label{eq:chooseansatz}
	\Gamma_{\kappa}[\tpsi,\psi] = \sum_x \int \! \mathrm{d}t\,  U_{\kappa}(\tpsi,\psi) + \action_{\epsilon} + \action_{Z = 1} \,,
\end{equation}
with the diffusion term  $\action_{\epsilon}$ and the term $\action_{Z = 1}$ as defined in Subsection~\ref{subsec:fieldtheoreticaction}. 
The initial condition for the effective average potential reads $U_{\kappa=\infty}(\tpsi,\psi) = \lambda(\mathbf{q}=\mathbf{0}) \tpsi^2 \psi^2 + \lambda(\mathbf{q}=\mathbf{0}) \tpsi \psi^2$, with $\lambda = \lambda(\bq=\mathbf{0}) = \sum_\bx \lambda(\bx)$.  
According to our above discussion, Eq.~(\ref{eq:chooseansatz}) includes all the terms needed to determine the correction exactly, even after subsituting
\begin{equation*}
U_{\kappa}(\tpsi,\psi)\to \mu \tpsi^2 \tpsi^2 + \mu \tpsi \psi^2 \,, \quad \epsilon(\mathbf{q}) \to q^2 \,.
\end{equation*} For the renormalized non-equilibrium force $F_{\kappa}(\psi) = \frac{\partial U_{\kappa}(\tpsi = 0,\psi)}{\partial \tpsi}$ we then obtain
\begin{equation}
\label{eq:fluct}
	\partial_{\kappa}  F_{\kappa}(\rho) \approx  \frac{2 \widetilde{V}_d  \kappa^{d+1} \mu^2 \rho^2}{\left(\kappa^2+2 \mu \rho \right)^2}\,,
\end{equation}	
where $\widetilde{V}_d$ stands for the volume of the $d$-dimensional sphere with radius $(2 \pi)^{-1}$. 
Although this equation is not exact, it still delivers the correct non-analytic contribution of the order $\rho^{\frac{d+2}{2}}$, because it treats all the terms which give rise to it exactly. 
Indeed, integrating Eq.~(\ref{eq:fluct}) from any $\kappa = \Lambda > 0$ to $\kappa=0$ yields a  contribution  
\begin{equation*}
	- \frac{\pi^{1-\frac{d}{2}}}{\Gamma\!\left(\frac{d}{2}\right) \sin\left(\frac{\pi d}{2}\right)}\left(\frac{\mu \rho}{2}\right)^{\frac{d}{2}+1}    \,,
\end{equation*}
(where  $\Gamma$  denotes the $\Gamma$-function and not the average action) which is valid for a dimension $d > 2$, as long as it is not an even natural number.  We notice that the correction alternates its sign, from a positive contribution in three dimensions, to a negative in five dimensions, and so on.  As $d$ approaches an even number, the result diverges, indicating logarithmic correction terms. In four and six dimensions, for instance, we find that a term
\begin{equation*}
	- \frac{\mu^3 \psi^3 \ln(\mu \rho)}{8 \pi^2} \,, \quad \frac{\mu^4 \psi^4 \ln(\mu \rho)}{32 \pi^3} \,,
\end{equation*}
respectively, is added to the force $F$. The $\kappa$-dependent logarithmic term in Eq.~(\ref{eq:alternatingseq}) indeed cancels. We remark that the correction terms are also amenable to perturbative treatment~\cite{Winkler:2012p23673, Note1}.

\begin{figure}
\centering
    \includegraphics[width=0.45\textwidth]{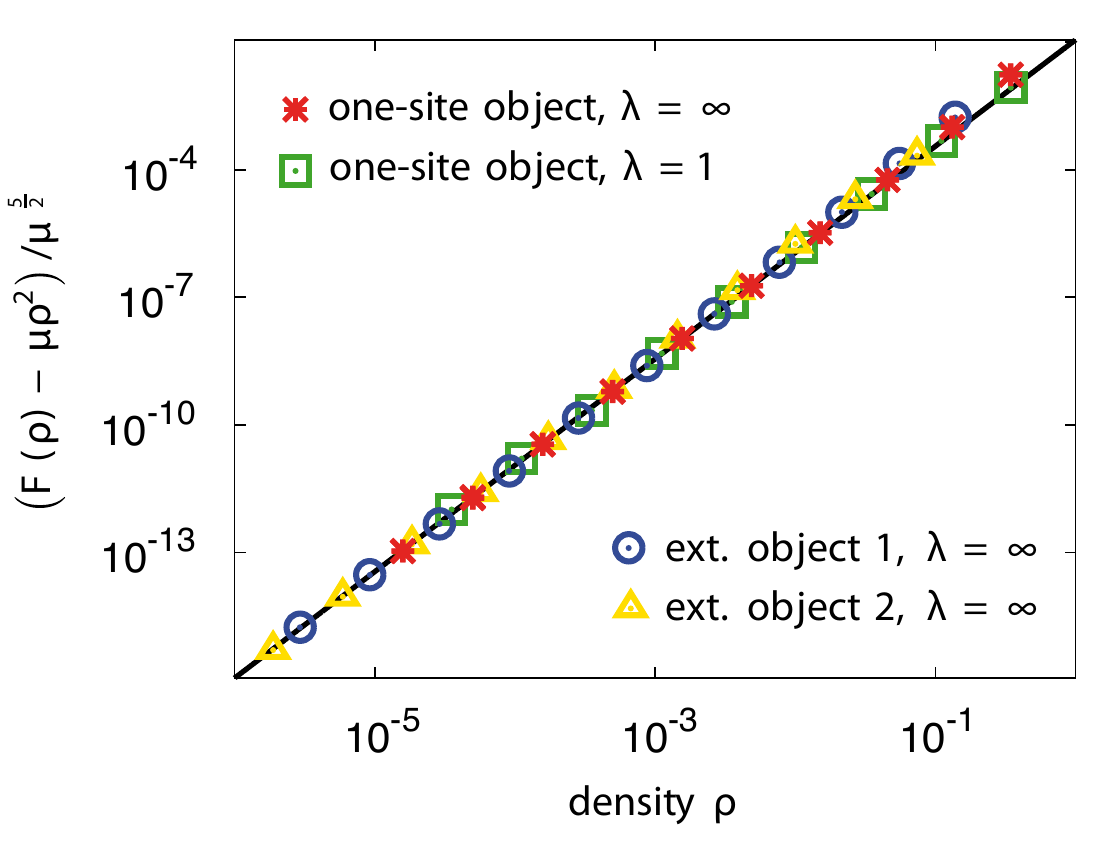}
    \caption{Rescaled data for the universal correction to the non-equilibrium force $F$. 
In the stochastic simulations, $F$ was determined directly by introducing homogeneous particle input and considering stationary states. On this double logarithmic plot, we show the rescaled data for $\left(F(\psi) - \mu \rho^2\right)/\mu^{\frac{5}{2}}$ for a range of models. We predict this term to be of the universal form  $\rho^{\frac{5}{2}}/\left(2 \sqrt{2} \pi\right)$ (solid black line), independent of the model, cf.~Eq.~(\ref{eq:correction}). The data evidently corroborate our theoretical results. 
 }
\label{fig:correction}
\end{figure}

We have run simulations for a  range of different models (one-site objects with both finite and infinitely large reaction rates, and two examples of extended objects that react immediately on contact) for the three-dimensional system, where 
\begin{equation}
\label{eq:correction}
	F(\rho) = \mu \rho^2 + \frac{\mu^{\frac{5}{2}}}{2 \sqrt{2} \pi}\rho^{\frac{5}{2}}  \,.
\end{equation}
The simulation results clearly corroborate our theoretical findings (see Fig.~\ref{fig:correction}).

\section{Conclusion}

In this article we presented results of our study of the coagulation process by means of a non-perturbative renormalization group (NPRG). Below the critical dimension the renormalization group flow drives the dimensionless non-equilibrium force $f_\tau(\chi)$ to a unique fixed point $f^\star(\chi)$. Within a certain approximation we can calculate $f^\star(\chi)$ and thus derive the ensuing anomalously slow, universal long-time density decay $\rho (t) \sim \mathcal{A}_d t^{-\frac{d}{2}}$. As the dimension is lowered from 2, the universal amplitude $\mathcal{A}_d$ becomes a non-perturbative quantity, beyond the reach of the perturbative approach. By considering the power law regime of  $f^\star(\chi)$ for large $\chi$, the NPRG enables us to extract the universal amplitude $\mathcal{A}_d$  even in one dimension where the estimate for $\mathcal{A}_d$ is in good agreement with exact calculations. 

Above the critical dimension, the long-time decay is governed by the law of mass action (LMA), such that $\partial_t \rho = - \mu \rho^2$ to lowest order in $\rho$, and $\rho (t) \sim \mu^{-1} t^{-1}$ in the long-time limit. NPRG provides a closed formula for the macroscopic decay rate $\mu$. 
Starting from the microscopic rate (which may be infinitely large) fluctuations in space and time are integrated gradually, going from short wavelengths and frequencies to long ones. Along the renormalization group flow the effective decay rate becomes monotonously smaller until all contributions are integrated and one obtains the macroscopic decay rate $\mu$. We work out the solution to the flow equation for the decay rate for a number of examples which allow for a particularly accurate solution.

Furthermore, we find that there are correction terms to the non-equilibrium force $F$ violating the law of mass action, which is generalized to $\partial_t \rho = - F(\rho) = - \mu \rho^2 + \ldots$ We identify a non-analytic term $c_d (\mu \rho)^{\frac{d+2}{2}}$ in the non-equilibrium force and show that it originates in long-range and short-frequency fluctuations. The term is universal in the sense that the amplitude $c_d \mu^{\frac{d+2}{2}}$ depends on the particular features of the process only through the non-universal rate $\mu$. The factor $c_d$ is completely independent of the microscopic details of the reaction process. For the three-dimensional case, we have run stochastic simulations which clearly confirm the theoretical predictions. 

The NPRG is a versatile and powerful tool for the study of non-equilibrium systems and the coagulation process is not only a simple model for the analysis of more complex systems but also serves as a starting point for the study of more complicated theoretical models~\cite{Canet:2004p83,*Canet:2004p18297,*Canet:2005p4749,Canet:2006p16432,Cardy:1996p279,*Cardy:1998p216,Benitez:2012p25246}.  We therefore expect that our results are relevant for a range of experimental systems and theoretical models
and believe they will encourage further work on the fundamental implications of fluctuations on non-equilibrium processes. 
The strong impact of fluctuation on the coagulation process below the critical dimension has already been probed by experimental studies on effectively one-dimensional exciton dynamics~\cite{BLAKLEY:1990p17131,Kroon:1993p15523}.
Similar experiments on excitons which disperse in all three spacial directions~\cite{Avakian:1968p17336} may be suitable for analysis of our predictions of a violation of the LMA in three dimensions. 

Financial support of Deutsche Forschungsgemeinschaft through the German Excellence Initiative via the program `Nanosystems Initiative Munich' (NIM) and through the SFB TR12 `Symmetries and Universalities in Mesoscopic Systems' is gratefully acknowledged.

\appendix*

\section{The Macroscopic Decay Rate for Selected Reaction Kernels}

\begin{figure}
      \centering \includegraphics[width=0.45\textwidth]{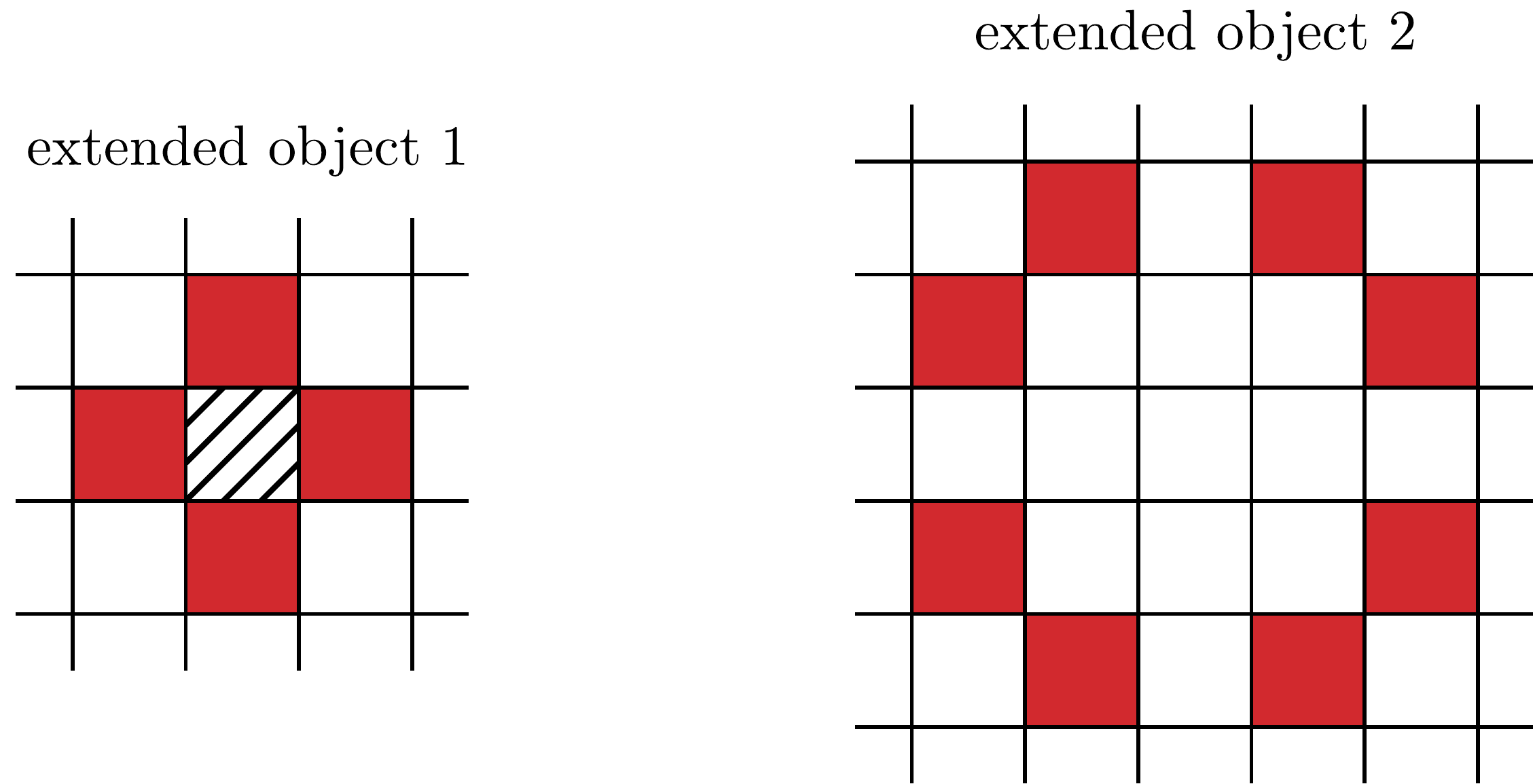}
\caption{Two-dimensional versions of the reaction kernels of extended objects (solid red). 
In three dimensions, for extended object 1, the kernel $\lambda(\bz) = \tilde \lambda_\infty$ if $\bz \in S = \{(\pm 1,0,0), (0,\pm 1,0), (0,0,\pm 1) \}$.  Otherwise it is zero. 
Notice that for instantaneous reactions, the striped square  can be regarded as part of extended object 1, which is then a discretization of the sphere. 
The support $S$ (with $\lambda(\bz) =  \tilde\lambda_\infty$ if $\bz \in S$) of the  three-dimensional reaction kernel of extended object 2 is crated by the union of the set $\{(1,1,2),(1,2,1),(2,1,1)\}$ with its mirror images in each octant. 
Since the flow conserves the support of the reaction kernel, their shape remains the same. 
For instantaneous coagulation reactions  we find $\mu^{-1} = 0.086064343192(3)$ (extended object 1) and $\mu^{-1} = 0.036287603611(2)$ (extended object 2). }
\label{fig:extended_objects}
\end{figure}

In the following we study reaction kernels which allow for a particularly precise numerical solution, repeating for completeness the calculations given in the Supplementary Material of~\cite{Winkler:2012p23673}. 
Consider the reaction kernels whose two-dimensional versions are depicted in Fig.~\ref{fig:extended_objects}. We study their three-dimensional versions on a cubic lattice. 
The lattice sites that these objects consist of (i.e. the sites that are part of the reaction kernel) are all equivalent. This symmetry must hold along the renormalization group flow. Furthermore, the flow equation conserves the support of the reaction kernel in position space. Thus the objects also retain their shape. 
Explicitly, in three dimensions the renormalized reaction kernel of extended object 1 can be expressed as
\begin{equation*}
	\lambda_k(\mathbf{p}) = \tilde\lambda_k \sum_{\nu = 1}^3 \left( e^{+ i p_\nu}  + e^{-i p_\nu}\right)  =  2 \tilde\lambda_k \sum_{\nu = 1}^3 \cos(p_\nu) \,,
\end{equation*}
The microscopic decay rate is $\lambda := \lambda_{\infty} (\mathbf{0}) = 6 \tilde\lambda_\infty$ ($\tilde \lambda_\infty$ is the reaction rate for each site of the kernel). Solving  Eq.~(\ref{eq:flow_lambda_p}) at $\mathbf{p}=\mathbf{0}$, one obtains the macroscopic decay rate
 \begin{IEEEeqnarray*}{C}
 	\frac{1}{\mu}  =  \frac{1}{\lambda_0(\mathbf{0})} = \frac{1}{\lambda_\infty(\mathbf{0})}  +  \\ + \frac{1}{18} \int_0^\infty \!\mathrm{d}k \int_{\mathbf{q}}  \left( 2 \sum_{\nu=1}^3 \cos(q_\nu) \right)^2 \frac{\Theta\!\left(k^2 - \epsilon(\mathbf{q}) \right)}{k^3}  =  
	\nonumber \vspace{0.5em} \\
	  =  \frac{1}{\lambda} + \underbrace{\int_{\mathbf{q}}  \frac{\left(\sum_{\nu = 1}^3 \cos(p_\nu)\right)^2}{36 \sum_{\nu=1}^3 \sin^2(q_\nu/2)}}_{0.086064343192(3)}\,.
\end{IEEEeqnarray*}
In the last step we inserted the dispersion relation $\epsilon(\mathbf{p}) = 4 \sum_{\nu=1}^3 \sin^2\left(\frac{p_\nu}{2}\right)$ for  the cubic lattice with unit lattice spacing.
The calculation for extended object 2 is analogous. If each of the 24 sites effects a reaction with rate $\tilde\lambda_\infty$ then
\begin{equation*}
	\frac{1}{\mu} = \frac{1}{\lambda} + 0.036287603611(2)\,,
\end{equation*}
with the microscopic decay rate $\lambda = 24 \tilde\lambda_\infty$. The simulations confirm these results for the macroscopic decay rates and support our prediction on the universal correction to the non-equilibrium force, cf.~Fig.~\ref{fig:correction}.

Finally we choose the reaction kernel to be the surface of a sphere, $\lambda_k(\mathbf{z}) = \tilde\lambda_k \, \delta(R - z)$, 
for objects diffusing in continuous space. 
In the limit of an infinitely large microscopic reaction rate this is evidently identical with spheres that coagulate instantaneously on contact.
Without loss of generality, we set the radius $R = 1$ in the following. Using spherical coordinates, the projection $\left( \mathcal{P} \circ \lambda_k\right) (\mathbf{z})$, see Eq.~(\ref{eq:projection}), becomes
\begin{IEEEeqnarray*}{C}
	  \frac{1}{(2 \pi)^3} \int \!\mathrm{d}q \ud\vartheta \ud\phi\, q^2\sin(\vartheta) e^{i q z \cos(\vartheta)} \Theta\!\left(k^2 - q^2\right) 
	\cdot  
	\nonumber \vspace{0.5em} \\
	 \cdot \:	\int \!\mathrm{d} r \ud \tilde\vartheta\ud\tilde\phi \, r^2 \sin(\tilde\vartheta) e^{-i q r \cos(\tilde\vartheta)} \tilde\lambda_k \,\delta(1-r) = 
	\nonumber \vspace{0.5em} \\
	  =  \tilde\lambda_k \underbrace{\frac{2}{\pi} \int_0^k \!\mathrm{d} q\, \frac{1}{z} \sin(q z) \sin(q)}_{=: f_k(z)}\,.
\end{IEEEeqnarray*}
Thus  from Eq.~(\ref{eq:flow_lambda_x}) we have $\partial_k \tilde\lambda_k  = 2 \tilde\lambda_k^2 \frac{f_k(1)}{k^3}$ and  
\[
 \frac{1}{\tilde\lambda_0}   =  \frac{1}{\tilde\lambda_\infty} + 2 \int_0^\infty \!\mathrm{d} k\, \frac{f_k(1)}{k^3} = \frac{1}{\tilde\lambda_\infty} +1\,.
\]
The macroscopic decay rate becomes
\begin{equation*}
	\frac{1}{\mu} = \frac{1}{\lambda} + \frac{1}{4\pi} \,, 
\end{equation*}
where $\lambda = 4\pi \tilde \lambda_\infty$. Therefore, for instantaneous reactions
\begin{equation*}
	\mu = 4 \pi D R\,,
\end{equation*}
with diffusion constant $D$ and radius $R$. This confirms Smoluchowski's result~\cite{Smoluchowski:1917p16756}, proved to be exact by Doi~\cite{Doi:1976p15932}.

\end{document}